%% file: main.tex
\documentclass[conference]{IEEEtran}
\IEEEoverridecommandlockouts
\ifCLASSINFOpdf
  \usepackage[pdftex]{graphicx}
\else
\fi
\usepackage{amsmath}
\usepackage{algorithmic}
\usepackage{amsmath,amsfonts}
\usepackage{algorithmic}
\usepackage{graphicx}
\usepackage{textcomp}
\usepackage{xcolor}
\usepackage{hyperref}
\usepackage{tikz}
\usepackage{amsmath}
\usepackage{multirow}
\usepackage{tabularx}
\usepackage{bbding}
\usepackage{caption}
\usepackage{subcaption}
\usepackage{pifont}
\usepackage{xspace}
\usepackage{booktabs} 
\usepackage{makecell}
\usepackage{tablefootnote}

\input{sections/commands}
\usepackage[normalem]{ulem}
\usepackage{xcolor}

\newcommand{\OLD}[1]{\iffalse #1\fi}
\newcommand{\NEW}[1]{{\noindent\color{black}{#1}}}

\begin{document}
%
\title{Incorporating Gradients to Rules: Towards Lightweight, Adaptive Provenance-based Intrusion Detection}

	


\author{\IEEEauthorblockN{Lingzhi Wang\IEEEauthorrefmark{1},
Xiangmin Shen\IEEEauthorrefmark{1},
Weijian Li\IEEEauthorrefmark{1}, 
Zhenyuan Li\IEEEauthorrefmark{3},
R. Sekar\IEEEauthorrefmark{4},
Han Liu\IEEEauthorrefmark{1}, and
Yan Chen\IEEEauthorrefmark{1}}
\IEEEauthorblockA{\IEEEauthorrefmark{1}Northwestern University, \IEEEauthorrefmark{3}Zhejiang University, \IEEEauthorrefmark{4}Stony Brook University}
\IEEEauthorblockA{\{lingzhiwang2025, xiangminshen2019, weijianli\}@u.northwestern.edu, \{hanliu, ychen\}@northwestern.edu\\
lizhenyuan@zju.edu.cn, sekar@cs.stonybrook.edu}
}



\maketitle

\pagestyle{plain}

\begin{abstract}
\input{sections/Abstract}
\end{abstract}


%
\IEEEpeerreviewmaketitle

\section{Introduction}
\label{sec:intro}
\input{sections/Introduction}

\section{Background and Related Work}
\label{sec:background}
\input{sections/Background}


\section{Motivation Examples}
\label{sec:motivation}
\input{sections/Motivation}

\section{System Design}
\label{sec:design}
\input{sections/Design}
\section{Implementation}
\label{sec:implementation}
\input{sections/Implementation}

\section{Evaluation}
\label{sec:evaluation}
\input{sections/Evaluation}

\section{Discussion and Future Work}
\label{sec:discussion}

\input{sections/Discussion}


\section{Conclusion}
\label{sec:conclusion}
\input{sections/Conclusion}


\section*{Acknowledgment}
We would like to thank anonymous reviewers for their constructive feedback. 
Lingzhi and Xiangmin were supported by the National Science Foundation (NSF) grant 2148177 and funds from the Resilient \& Intelligent NextG Systems (RINGS) program.
In addition, Zhenyuan was supported by the ``Pioneer" and ``Leading Goose" R\&D Program of Zhejiang (2024C03288).
Sekar's work was supported in part by NSF grants 1918667 and 2153056.



\bibliographystyle{IEEEtran}
\bibliography{refs}

%



\appendices
\label{sec:appendix}
\input{sections/Appendix.tex}

\end{document}

%% file: sections/commands.tex
\newcommand{\SysName}{{\scshape Captain}\xspace}

\newcommand{\MORSE}{{\scshape Morse}}
\newcommand{\NODOZE}{{\scshape NoDoze}}
\newcommand{\SLEUTH}{{\scshape Sleuth}}
\newcommand{\CONAN}{{\scshape Conan}}
\newcommand{\SHADEWATCHER}{{\scshape ShadeWatcher}}
\newcommand{\UNICORN}{{\scshape Unicorn}}
\newcommand{\HOLMES}{{\scshape Holmes}}
\newcommand{\PROVDETECTOR}{{\scshape ProvDetector}}
\newcommand{\WATSON}{{\scshape Watson}}
\newcommand{\CADETS}{{\scshape Cadets}}
\newcommand{\TRACE}{{\scshape Trace}}
\newcommand{\THEIA}{{\scshape Theia}}
\newcommand{\ATLAS}{{\scshape Atlas}}
\newcommand{\POIROT}{{\scshape Poirot}}
\newcommand{\FLASH}{{\scshape Flash}}
\newcommand{\KAIROS}{{\scshape Kairos}}
\newcommand{\PROGRAPHER}{{\scshape ProGrapher}}
\newcommand{\NODLINK}{{\scshape NodLink}}
\newcommand{\THREATRACE}{{\scshape ThreaTrace}}
\newcommand{\RCAID}{{\scshape R-caid}}

\newcommand{\codeRepo}{https://github.com/LexusWang/CAPTAIN}

%% file: sections/Abstract.tex
As cyber attacks grow increasingly sophisticated and stealthy, it becomes more imperative and challenging to detect intrusion from normal behaviors.
Through fine-grained causality analysis, provenance-based intrusion detection systems (PIDS) demonstrated a promising capacity to distinguish benign and malicious behaviors, attracting widespread attention from both industry and academia.
Among diverse approaches, rule-based PIDS stands out due to its lightweight overhead, real-time capabilities, and explainability. 
However, existing rule-based systems suffer low detection accuracy, especially the high false alarms, due to the lack of fine-grained rules and environment-specific configurations.

In this paper, we propose \SysName, a rule-based PIDS capable of automatically adapting to diverse environments.
Specifically, we propose three adaptive parameters to adjust the detection configuration with respect to nodes, edges, and alarm generation thresholds.
We build a differentiable tag propagation framework and utilize the gradient descent algorithm to optimize these adaptive parameters based on the training data.
We evaluate our system using data from DARPA Engagements and simulated environments.
The evaluation results demonstrate that \SysName enhances rule-based PIDS with learning capabilities, resulting in improved detection accuracy, reduced detection latency, lower runtime overhead, and more interpretable detection procedures and results compared to the state-of-the-art (SOTA) PIDS.\newline

\noindent Note: This is a preprint version of the paper accepted
at the Network and Distributed System Security Symposium (NDSS) 2025.

%% file: sections/Introduction.tex


Advanced Persistent Threats (APT) are becoming a growing threat to both government and industrial sectors, causing significant societal impacts~\cite{apt_survey}.
The Equifax breach in 2017 resulted in the theft of vast amounts of personal data, highlighting the severe privacy and security risks~\cite{noauthor_target_nodate, noauthor_equifax_nodate}.
Moreover, attackers continuously innovate to find new ways to penetrate systems and remain undetected for extended periods.
In recent years, provenance-based intrusion detection systems (PIDS) have gained attention from both the security industry and academia for their causality analysis capability.
However, prevailing alarm fatigue, excessive runtime overhead, long detection latency, and opaque detection processes (based on black-box) are still open research problems in real-world scenarios~\cite{dong2023we, fuzzy_fatigue, anomaly_survey, nodoze, shen2024decoding}.

Alarm fatigue~\cite{soc_perspective, sundaramurthy2015human} is a significant issue plaguing the security industry.
A recent survey~\cite{99FP} indicates that security analysts are required to handle an overwhelming average of 5,000 alarms daily, with a majority being false alarms.
The excessive volume of alarms can lead to serious consequences.
For instance, in the 2013 Target data breach~\cite{noauthor_target_nodate}, the malicious activities were detected and reported by security tools but overlooked by analysts, resulting in delayed response and expanded losses~\cite{target_report}.
In practice, false alarms are equally detrimental as missed alarms. 
Moreover, significant challenges also arise from runtime overhead and detection latency. High runtime overhead compromises the scalability of detection systems, hindering their deployment on a large scale~\cite{dong2023we}. Moreover, detection latency critically affects performance, as prolonged latency delays the response to threats, thereby impeding analysts' efficiency in managing alarms.


In academia, a recent trend of PIDS~\cite{zengy2022shadewatcher, yang2023prographer, cheng2023kairos, rehman2024flash} leverage embedding techniques like word2vec~\cite{rehman2024flash} and graph2vec~\cite{yang2023prographer} to encode system entities and events, and neural networks like Graph Neural Networks (GNNs)~\cite{rehman2024flash, cheng2023kairos, zengy2022shadewatcher,goyal2024r} and Recurrent Convolutional Neural Networks (RCNNs)~\cite{yang2023prographer} to analyze information flow and produce detection results. We refer to them as \textit{embedding-based PIDS}.
Although these systems have demonstrated notable achievements in detection performance, they face the following challenges: 
\begin{list}{\labelitemi}{\leftmargin=3pt}
 \setlength{\itemsep}{3pt}
 \setlength{\itemindent}{9pt}
    \item \textbf{High Computational Resource Cost.} Embedding-based systems need to embed the graph features into vectors, usually with deep learning techniques~\cite{yang2023prographer, rehman2024flash, zengy2022shadewatcher, cheng2023kairos}, which require a significant amount of computational resources.
    Moreover, many systems require caching of embeddings and deep learning models, leading to high memory consumption~\cite{dong2023we};
    \item \textbf{Long Detection Latency.} 
    Embedding-based systems take the input structured as graphs or paths, requiring time windows~\cite{cheng2023kairos,li2023nodlink} or log batches~\cite{rehman2024flash, yang2023prographer} in their design, which leads to the additional detection delay.
    \item \textbf{Uninterpretable Results.} Many systems~\cite{han2020unicorn, zengy2022shadewatcher, wang2022threatrace,rehman2024flash,cheng2023kairos,li2023nodlink} only flag deviations from normal behaviors without attack semantics.
    Moreover, we are unable to open the ``black box" of the deep learning models to better understand the detection process.
\end{list}


On the other hand, a group of PIDS~\cite{hossain2017sleuth, morse, holmes, conan, nodoze} operates based on human-defined mechanisms to generate representations for system entities, propagate information flow, and trigger alarms. 
We refer to them as \textit{rule-based PIDS}.
Leveraging these rules, they take actionable leads, allowing themselves to provide semantic-rich alarms and pinpoint specific suspicious events for further investigation. 
Besides, the rule-based approach operates on simple arithmetic calculations, which require far fewer computational resources than complex matrix calculations by neural networks.
Furthermore, some rule-based approaches process streaming data once it arrives, significantly reducing storage requirements and enabling rapid response with minimal delay.
In summary, rule-based PIDS shows advantages in real-world detection tasks due to its fine-grained detection, semantic-rich alarms, lightweight overhead, and minimal detection delay.

Despite rule-based PIDS excelling in the aforementioned perspectives, they also face challenges in real-world deployment.
These systems often employ simplistic and universal rules to distinguish between graphs (such as discrete trustworthiness levels of node in \cite{morse,hossain2017sleuth,holmes}, universal system parameters in \cite{morse,hossain2017sleuth,holmes,nodoze}). The inflexible rules lead these PIDS to be either too lenient or too strict.
In security operations centers (SOCs), analysts have to manually configure the models, which is a time-consuming process~\cite{kokulu2019matched}.
This highlights the need for an automated configuration methodology and scheme. 
To achieve this, a rule-based PIDS must be capable of adjusting its rules autonomously based on detection feedback from the training set, without human intervention.
This requires an automatic feedback mechanism that establishes a direct connection between the rules and detection results, allowing the system to dynamically adjust its configuration based on the detection outcomes.
In this paper, we propose \SysName, a rule-based PIDS with adaptive configuration learning capability.
We aim to leverage the advantages of traditional rule-based PIDS while enabling the system to acquire suitable configurations autonomously.
Specifically, we introduce three adaptive parameters to the rule-based PIDS, giving it more flexibility during detection.
Moreover, we design a learning module to adjust these parameters automatically based on the detection results during training.
We calculate and record the gradients of each adaptive parameter, transferring the rule-based system to a differentiable function.
With the help of a loss function, the gradient descent algorithm can be utilized to optimize the adaptive parameters based on the benign training data and thus reduce the false alarms in testing.

We evaluate \SysName in diverse detection scenarios, which are drawn from the widely acknowledged public datasets~\cite{apt_note} provided by DARPA Engagement~\cite{noauthor_transparent_nodate} and datasets collected from simulated environments in collaboration with a SOC.
The evaluation results demonstrate that \SysName can reduce false alarms by over 90\% (11.49x) on average compared to traditional rule-based PIDS.
Additionally, \SysName achieves much better detection accuracy with less than 10\% CPU usage and significantly lower memory usage and latency compared to the SOTA embedding-based PIDS.
We then select a few scenarios as cases to study the explainability of the learned configurations.

In summary, this paper makes the following contributions: 

\begin{list}{\labelitemi}{\leftmargin=3pt}
 \setlength{\itemsep}{3pt}
 \setlength{\itemindent}{9pt}
\item We propose \SysName, a rule-based PIDS that can adjust its rules using benign data to reduce false alarms.
Our design endows \SysName with the lightweight nature, low latency, and interpretability of rule-based PIDS while also having fine detection capability and adaptability.

\item We parameterize the rule-based PIDS and propose a novel differentiable tag propagation framework that allows us to optimize adaptive parameters using the gradient descent algorithm.

\item We systemically evaluate \SysName across various scenarios, including widely used DARPA datasets and simulated datasets from a partner SOC.
The evaluation result demonstrates that \SysName can automatically adapt to various environments, offering superior detection accuracy with reduced latency, lower overhead, and more interpretable detection.
\end{list}

We make the code of our system, all datasets, and all experiments publicly available for future analysis and research\footnote{\codeRepo}.

%% file: sections/Background.tex
In this section, we begin by introducing PIDS and highlighting mainstream approaches to clarify our design choices.
Next, we provide background on the optimization problem, followed by a description of the threat model and the assumptions.

\subsection{Provenance-based Intrusion Detection}
\label{subsec:background-PIDS}
Provenance-based intrusion detection has drawn wide attention for its powerful correlation and causal analysis capabilities. 
In provenance analysis, the system behaviors are modeled as directed acyclic graphs, called provenance graphs~\cite{zipperle,li2021threat}, in which nodes represent system entities, such as processes, files, sockets, pipes, memory objects, etc., and edges represent interactions between entities, such as reading files and connecting to a remote host. 

\begin{figure}[htbp]
\vspace{-0.5em}
    \centering
    \includegraphics[width=1.0\linewidth]{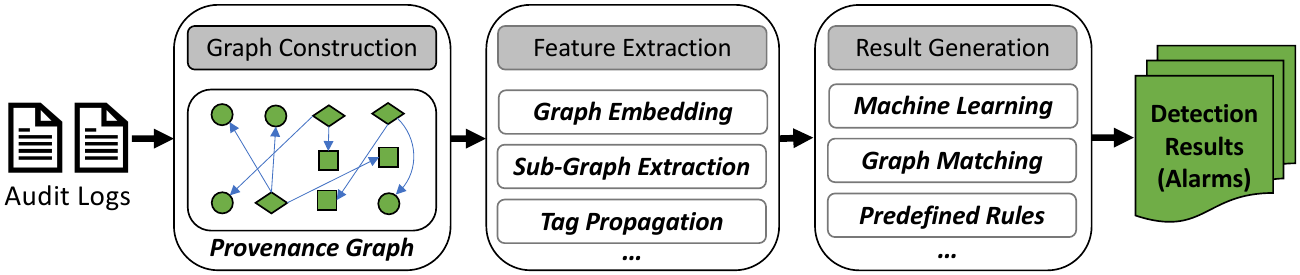}
    \caption{A brief overview of the workflow and commonly used techniques in mainstream PIDS.}
    \label{fig:principles_overview}
    \vspace{-0.5em}
\end{figure}

Fig.~\ref{fig:principles_overview} illustrates an overview of typical PIDS workflow.
Based on how to model and aggregate the entity information, the data/control flow, and the system behaviors, we identify two major groups of PIDS: the embedding-based approach and the rule-based approach.

\noindent\textbf{Embedding-based PIDS.}
Embedding-based systems~\cite{li2023nodlink, han2020unicorn, zeng2021watson, alsaheel2021atlas, zengy2022shadewatcher,rehman2024flash,cheng2023kairos,goyal2024r} use numerical vectors to represent, propagate, and aggregate the information of system entities, events, data/control flow in the graph.
Various techniques in machine learning have been employed to embed the graph features into numerical vectors.
Since these techniques typically require the input structured as graphs or paths, embedding-based PIDS usually adopt time windows or log batches in their design, leading to the additional detection delay, which we named as \textit{\textbf{buffer time}} in this paper.
Concretely, buffer time refers to the time window duration in which all streaming logs are stored to create a graph structure for subsequent processing.
Previous studies have implemented a fixed-length window based on time~\cite{cheng2023kairos} or the number of events~\cite{rehman2024flash}, while some works~\cite{yang2023prographer} utilize the variable length window.
Buffer time is crucial in real-world detection scenarios because, in the worst-case scenario, after an attack has been launched, the PIDS would have to wait for the duration of the buffer time before starting the detection procedure.

The methods for embedding graphs differ across various systems.
For instance, \PROVDETECTOR~\cite{wang2020you} selects rare paths based on historical event frequency to generate embedding vectors for sequence learning and anomaly detection.
\ATLAS~\cite{alsaheel2021atlas} classifies the extracted sequences from the graph with Long Short-term Memory (LSTM).
\UNICORN \cite{han2020unicorn} embeds the graph histograms and clusters the embedded graph sketches.
\WATSON~\cite{zeng2021watson} learns the node embedding and event semantics from training traces.
More recent embedding-based systems~\cite{rehman2024flash, cheng2023kairos,goyal2024r} employed graph neural networks in detection.
While such systems can achieve better detection accuracy, they fail to provide the rationale for generated alarms. 
For example, \SHADEWATCHER~\cite{zengy2022shadewatcher} provides the probability of malicious system events without further context. \FLASH~\cite{rehman2024flash} highlights suspicious entities with causal links but does not provide further attack semantics of those interactions.
In practice, more manual efforts are required to review the relevant audit logs thoroughly for further response.

\noindent\textbf{Rule-based PIDS.}
Rule-based systems~\cite{hossain2017sleuth, conan, holmes, morse, nodoze} leverage human-designed rules to map system entities and events to predefined semantic units (e.g., tags~\cite{hossain2017sleuth, morse, conan}, TTP~\cite{holmes}, etc.) and propagation these units along the information flow to capture causality.
\NODOZE \cite{nodoze} assigns an anomaly score to each edge based on the frequency it appears in the benign data and propagates the scores on the graph.
\POIROT~\cite{poirot} models the detection as a graph-matching problem that aligns manually constructed query graphs on the provenance graph.
\SLEUTH~\cite{hossain2017sleuth} and \MORSE~\cite{morse} assign tags to system entities and propagate tags among the graph.
Rule-based systems have become popular due to their real-time processing capabilities and lightweight properties. 
However, they usually suffer low detection accuracy due to the lack of fine-grained rules and environment-specific configurations.
Most rule-based PIDS~\cite{morse, hossain2017sleuth,conan,holmes} implement simple and universal rules, making them inflexible to adapt to different environments. 
These systems typically rely on a discrete classification to denote system entity properties, such as private/public, and trusted/untrusted, and apply identical rules to all nodes and events.
As a result, they lack the nuance required to accurately distinguish between similar graphs, often being either overly lenient or overly strict, which leads to false alarms or missed true alarms.
We will elaborate on these challenges in \S\ref{sec:motivation}.



Table \ref{table:pids-comparison} summarizes the detection granularity of existing PIDS and indicates whether the detection requires buffer time.


In this paper, we build our system \SysName following the rule-based methodology for four reasons:
1) rule-based detection is computationally more efficient~\cite{poirot, han2020unicorn, hossain2017sleuth};
2) rule-based methods are suitable for real-time detection because they can process the event stream incrementally~\cite{li2021threat};
3) rule-based methods offer more explainability to the detection process and results than embedding-based methods~\cite{welter2023tell};
4) rule-based methods are robust against mimicry attacks~\cite{goyalsometimes}.
Additionally, \SysName introduces a significant advancement over traditional rule-based PIDS.
Unlike systems that rely on simple, universal, and inflexible rules, \SysName offers a mathematically complete framework for automatically fine-tuning detection rules, enabling more precise detection and response to security threats.

                 

\begin{table}[h!]\footnotesize
\centering
\caption{Comparison of existing PIDS. Buffer time refers to the waiting time before threat analysis.}
\begin{tabular}{c|c|c}
\toprule
                 & \makecell{Detection\\Granularity} &  \makecell{Require\\Buffer Time} \\ \midrule
                 
\SysName        &  Edge     &  No   \\ \hline
\FLASH          &  Node     &   Yes  \\ \hline
\NEW{\RCAID}         &  \NEW{Node}   &   \NEW{Yes}  \\ \hline
\KAIROS         &  Graph$^1$   &   Yes  \\ \hline
\NODLINK        &  Node    &  Yes  \\ \hline
\PROGRAPHER     &  Node    &  Yes  \\ \hline
\SHADEWATCHER   &  Edge    &  Yes  \\ \hline
\POIROT         &  Path    &  Yes \\ \hline
\MORSE          &  Edge    &  No \\ \hline
\UNICORN        &  Graph   &  Yes  \\ \hline
\HOLMES         &  Edge    &  No  \\ \hline
\PROVDETECTOR   &  Path    &   Yes  \\ \hline
\NODOZE         &  Path    &   Yes   \\
\bottomrule
\end{tabular}
\\ \footnotesize \raggedright
$^1$Although \KAIROS\ triggers alarms on the graph level, it can highlight anomaly nodes and edges in the subgraphs.
\label{table:pids-comparison}
\vspace{-1.5em}
\end{table}

\subsection{Configuration as an Optimization Problem}
Finding the best configuration for a rule-based PIDS can be formalized as an optimization problem, which, in short, aims to find the best elements $\theta$ in a searching space to minimize or maximize an objective function $J(\theta)$.
Gradient descent is one of the most common algorithms to solve the optimization problem~\cite{ruder2016overview}.
It utilizes the gradient of the objective function to the parameters $\nabla_{\theta}J(\theta)$ to update $\theta$ according to $\theta = \theta - l \cdot \nabla_{\theta}J(\theta)$, where $l$ is the learning rate.
Many variants have been proposed based on the gradient descent algorithm.
We discuss more algorithms and their applicability to \SysName in~\S~\ref{sec:more_learning_algo}.

While the optimization problem and its gradient-based solutions play a significant role in quantitative science and engineering~\cite{martins2021engineering}, it is rarely discussed in the previous rule-based PIDS because of the following challenges.
First, no existing work formally defined the adjustable parameters with clear meanings in the context of rule-based PIDS.
Second, designing an objective function, calculating, updating, and saving the gradients needed in the gradient descent algorithm are knotty problems for a rule-based PIDS.

\subsection{Threat Model \& Assumptions}
\label{subsec:threatmodel}


Similar to many previous works \cite{morse,hossain2017sleuth,han2020unicorn,wang2020you,nodoze,li2023nodlink}, we assume that OS kernels and auditing tools are a part of the trusted computing base (TCB), which means the attackers are not able to tamper system auditing data.
Meanwhile, we do not consider hardware manipulation and any other attacks that leave no traces on the provenance graph.

In each detection environment, we presume the availability of audit logs encompassing daily activities and devoid of any malicious activities to serve as the training data.
In other words, we adopt the assumption that any alarm generated from this training data is a false alarm.







%% file: sections/Motivation.tex
\begin{figure}
    \centering
    \includegraphics[width=\linewidth]{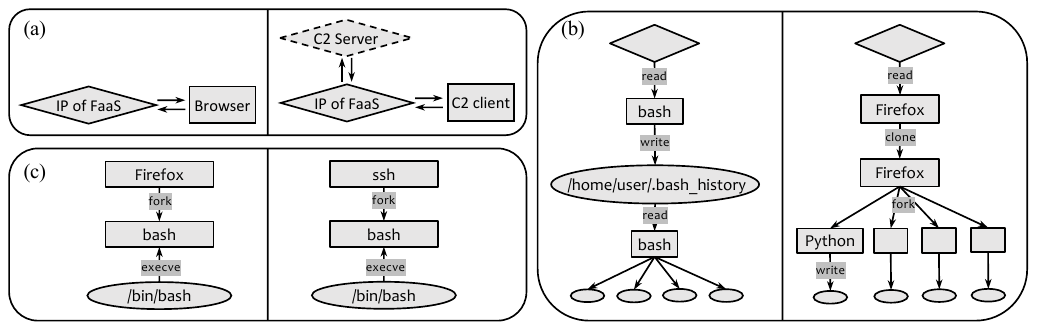}
    \caption{\NEW{Three motivating examples from the real-world dataset where the more fine-grained rules are needed in the rule-based PIDS.}}
    \label{fig:motivating-example}
    \vspace{-1.5em}
\end{figure}

In this section, we present motivating examples to elucidate the challenges encountered by traditional, coarse-grained rule-based PIDS, as depicted in Fig.~\ref{fig:motivating-example}.
These scenarios contain similar benign and malicious behaviors, which pose challenges to existing rule-based PIDS~\cite{morse,hossain2017sleuth,holmes,nodoze} in adjusting their rules to effectively select an optimal trade-off point between excessive false alarms and missing true alarms.

\textbf{``Grey" Nodes:}
With the advancement of cloud computing, connections to Function as a Service (FaaS) platforms like AWS Lambda and Cloudflare Workers have become a typical pattern in provenance graphs.
However, recent reports indicate that attackers are exploiting these commercial FaaS services to redirect traffic to Command and Control (C2) servers, as illustrated in Fig.~\ref{fig:motivating-example}a. 
Existing rule-based PIDS are often too coarse to depict those IPs.
For example, \SLEUTH~\cite{hossain2017sleuth} only defined three trustworthiness levels: \textit{Benign Authentic}, \textit{Benign}, and \textit{Unknown} for the nodes, and \MORSE~\cite{morse} used a binary initial tag for the trustworthiness.
But in this scenario, since both benign and malicious behavior can involve communications with the external FaaS IPs, the coarse-grained rules either lead to excessive false alarms or missing true alarms.
A more fine-grained configuration to reflect the ``grey scale" of these IPs may help solve the issue, i.e., we can assign a trustworthiness value (a float between 0 and 1) to every node in the provenance graph.

\textbf{Dependency Explosion:}
Dependency explosion refers to the scenario where all the subsequent events are treated as they depend on the previous ones, and thus a single suspicious event can lead to millions of system entities considered suspicious~\cite{morse}.
As shown in Fig.~\ref{fig:motivating-example}b, a prolonged and furcate event chain can spread the maliciousness carried by the suspicious entities to entities involved in all subsequent events.
Many existing efforts have been made to solve this issue~\cite{newsome2005dynamic, xu2006taint, arzt2014flowdroid, ma2016protracer,lee2013high}.
Unfortunately, their solutions usually require extensive instrumentation of applications/OS~\cite{arzt2014flowdroid, ma2016protracer,lee2013high} and expertise, which limits the usage in the real-world system, especially in end-point host deployment.

To deal with the dependency explosion problem without extensive instrumentation, existing rule-based PIDS adopt methods like event whitelisting~\cite{hossain2017sleuth}, cost-based pruning, or dividing suspicious events into different stages in a typical APT lifecycle~\cite{holmes}.
\MORSE~\cite{morse} proposed the decay and attenuation mechanisms to reduce the maliciousness carried by the system entities over time to mitigate the dependency explosion.
However, the decay and attenuation factor is the same for all nodes and edges.
Such identical configuration can lead to the dilemma between excessive false alarms and missing true alarms as attackers may deliberately fork many irrelevant processes or connect to benign entities to evade detection.
As shown by the two provenance graphs in Fig.~\ref{fig:motivating-example}b, we want to confine the dependency explosion in the left graph while preserving the maliciousness in the right (the right graph depicts a rare pattern where a \texttt{firefox} process forks a \texttt{Python} process, indicating the possibility of suspicious script execution from a website.)
Ideally, we aim to set a propagation rate for each individual event, allowing for more customized control.
This ensures that the maliciousness of unusual events does not fade away too quickly, while it can reset rapidly for more common events.

\textbf{Customized Alarm Triggering:}
In rule-based PIDS, alarms are triggered based on predefined conditions that are uniformly applied to all events, regardless of their context. 
These one-fits-all approaches can lead to suboptimal detection performance.
For instance, \MORSE\ sets a series of detection rules and thresholds to trigger alarms and the rules are identical for all processes and events.
As the motivating example shows in Fig. \ref{fig:motivating-example}c, \texttt{firefox} and \texttt{sshd} are two processes that often have network activities with various IP addresses. 
While \texttt{sshd} may commonly use \texttt{bash} for script execution, it is unusual and suspicious for \texttt{firefox} to do the same. Despite the similarities in event types and related entities between \texttt{firefox} and \texttt{sshd}, applying a uniform threshold for all ``fork" events could either result in false alarms for \texttt{sshd} or miss genuine threats involving \texttt{firefox}.
To distinguish between such two behaviors, a specific rule is needed: for the common events, we want to set a higher threshold to avoid excessive false alarms, while for rare events, we want to keep the threshold low to avoid missing true alarms.
A knottier question then follows: how to set different thresholds about the ``execute bash" event for each process?

The three motivating examples underscore the challenges that existing rule-based PIDS face in balancing false and true alarms due to their coarse-grained rules. Additionally, these systems lack the flexibility to dynamically adjust rules to accommodate different environments.
Theoretically, embedding-based PIDS, leveraging the power of deep learning, can effectively create customized ``rules" for every event and entity in the graph by embedding them into the feature space.
However, training and deploying deep neural networks require numerous calculation resources and storage space, leading to high runtime overhead.
Moreover, these PIDS usually lack the explainability of the detection process and result, making it difficult to pinpoint the malicious behaviors at the event level.

To overcome these limitations, we aim to refine rule-based PIDS using fine-grained rules.
By allowing the rule-based detector to learn and adapt from training data, we can enhance detection accuracy without the extensive resource demands and runtime overhead.
Our system improves the overall efficacy of rule-based PIDS but also maintains its operational simplicity and clarity.

\OLD{
\noindent\textbf{Entity-level Distinction:} 
Entity-level information plays a pivotal role in understanding system behaviors through provenance graphs.
System entities span various categories, such as processes, files, sockets, etc.
Each system entity is characterized by specific features: processes are identified by their names and command lines, files by their file paths, and sockets by their IP addresses and port numbers.}

\OLD{Embedding-based PIDS usually use machine learning techniques, especially from the natural language processing (NLP) field, to embed the entity-level features, such as hierarchical feature hashing~\cite{cheng2023kairos}, Word2Vec~\cite{rehman2024flash}, and FastText~\cite{li2023nodlink}.
However, these techniques are computationally expensive and inexplicable.
On the other hand, it is hard for rule-based PIDS to distinguish the system entities for the following reasons: 1. the amount of entities is huge; 2. the detailed entity-level configuration rules vary across different environments.
Therefore, in most cases, they have to apply general rules.
Unfortunately, such generic configurations often result in excessive false alarms.
As the motivating example in Fig. \ref{fig:motivating-example}a shows, if the detection systems do not distinguish the network nodes by their IP addresses, all system entities interacting with \texttt{sshd} would inherit maliciousness from those external connections and trigger false alarms.
}

\OLD{\noindent \textbf{Event-level Distinction:}
Event-level distinction is also challenging for the rule-based PIDS during detection.
System events serve as bridges that propagate the contextual information flows among system entities and form the kill chains for the attack.
However, traditional rule-based PIDS find it challenging to adaptively generate fine-grained propagation and detection rules for different environments.
Generic rules might lead to unsatisfying detection performance.
In our motivating example in Fig. \ref{fig:motivating-example}(c), \texttt{firefox} and \texttt{sshd} are two processes that often have network activities with various IP addresses. However, \texttt{sshd} is more likely to use \texttt{bash} to execute scripts, while it is suspicious for \texttt{firefox} to perform such tasks.
Please note that in this example, the related entities and event types are the same for \texttt{firefox} and \texttt{sshd}.
To distinguish such two behaviors, a very explicit rule is needed (A knotty question would be: do we need to set different rules about the ``execute bash" event for each different process?).
Some rule-based PIDS used frequency information in the training set to generate rules~\cite{nodoze}.
We will show later that their methods are more susceptible to the training set poisoning attack.
}

\OLD{\noindent\textbf{Dependency Explosion:}
Besides distinguishing system entities and events, dependency explosion is a common problem in provenance analysis~\cite{morse}.
In our motivating example in Fig. \ref{fig:motivating-example}(b), we observe that the \texttt{bash} processes frequently write to and read from the \texttt{.bash\_history} file.
Notably, the \texttt{bash} process interacts with several potentially malicious entities as demonstrated in Fig. \ref{fig:motivating-example}(a) and (c).
Therefore, the \texttt{bash} processes tend to have relatively high maliciousness in these scenarios.
Consequently, these \texttt{bash} processes propagate the maliciousness to the \texttt{.bash\_history} file and then \texttt{.bash\_history} spread the maliciousness among other system entities via subsequent interactions, triggering more false alarms.
Many existing efforts have been made to solve this issue~\cite{newsome2005dynamic, xu2006taint, arzt2014flowdroid, ma2016protracer,lee2013high}.
Unfortunately, their solutions usually require extensive instrumentation of applications/OS~\cite{arzt2014flowdroid, ma2016protracer,lee2013high} and expertise, which limits the usage in the real-world system, especially in end-point host deployment.
In this paper, we aim to address the dependency explosion issue via automatic learning without instrumentation on the auditing module. 
}

%% file: sections/Design.tex

\begin{figure}
    \centering
    \includegraphics[width=\linewidth]{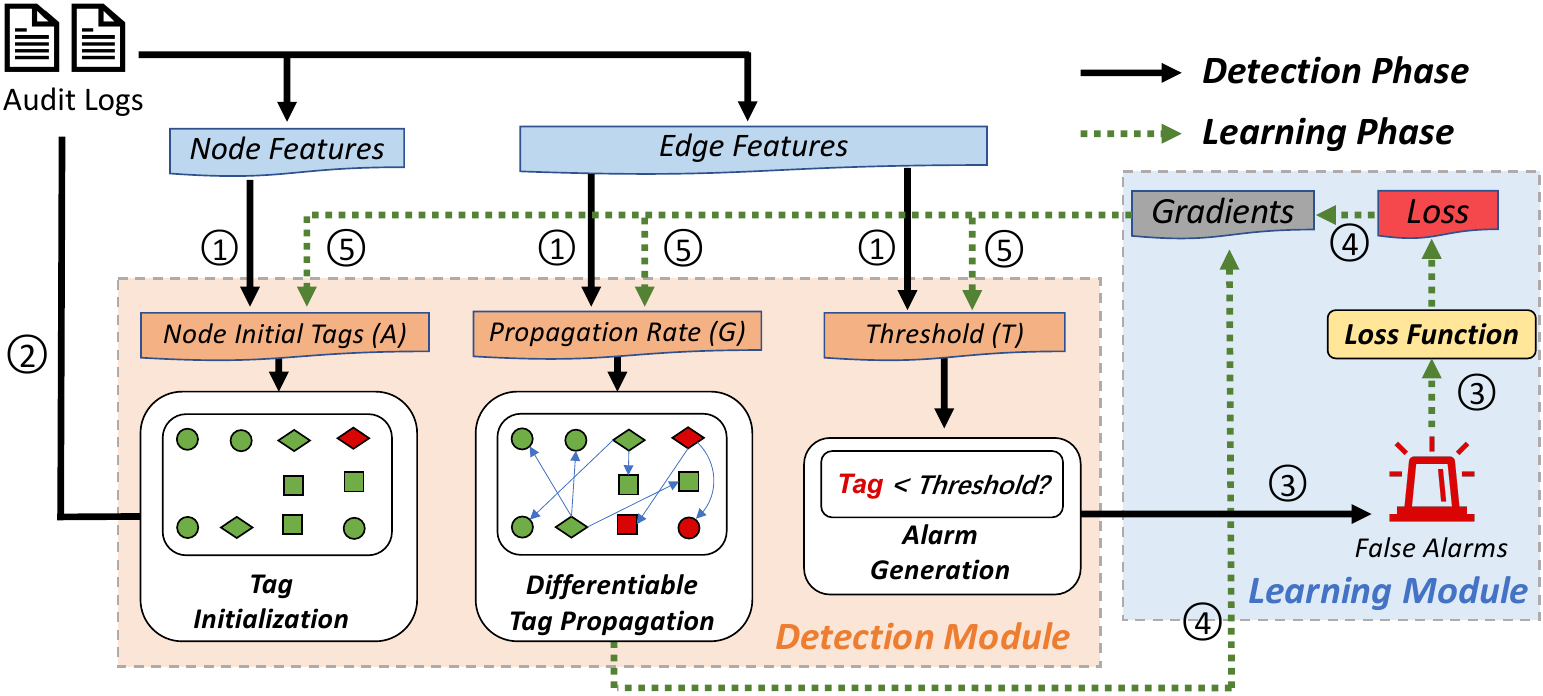}
    \caption{The overall framework of \SysName. Phases \normalsize{\textcircled{\scriptsize{1}}}-\normalsize{\textcircled{\scriptsize{5}}}\ show the lifecycle of the detection module and the learning module within one training epoch. 
    }
    \label{fig:overall-framework}
    \vspace{-1em}
\end{figure}

\subsection{System Overview}\label{sec:overalldesign}

As shown in Fig.~\ref{fig:overall-framework}, \SysName consists of two major parts: the detection module and the learning module.
The detection module takes the audit logs as input and produces detection results, while the learning module leverages the false alarms during the training phase to fine-tune the detection module.




\subsection{Adaptive Detection Module}\label{sec:detection_design}
The detection module in \SysName comprises three major components: tag initialization, tag propagation, and alarm generation.
System entities are assigned initial tags when they appear in the system.
As system events happen, tags get updated among system entities, which we refer to as ``tag propagation".
At the same time, \SysName determines whether to generate alarms based on the pre-defined rules.

\noindent\textbf{Tag Design:}
Inspired by previous provenance-based methodology based on data flow and control flow~\cite{park2012provenance,pasquier2017practical,hossain2017sleuth,morse}, \SysName designs two types of tags: data tags and code tags.
As shown in Table~\ref{tab:tags}, data tags exist on all nodes in the provenance graph.
A data tag is represented as a numerical vector \texttt{<c, i>}, where \texttt{c} denotes the confidentiality score of the data, and \texttt{i} denotes the integrity score of the data.
In contrast, code tags only exist on process nodes, denoted by \texttt{<p>} to indicate the integrity of the code.
All scores (\texttt{c}, \texttt{i}, and \texttt{p}) are real numbers ranging from 0 to 1.
A low integrity score suggests that the data/code might be compromised and, therefore, is considered untrusted.
A low confidentiality score is counterintuitively associated with highly sensitive data (to make the scoring scheme consistent), such as passwords.

\begin{table}[h!]\scriptsize
\centering
\caption{Tags in \SysName}
\label{tab:tags}
\begin{tabular}{ll|l}
\toprule
\multicolumn{2}{l|}{Tag Type}                                        & Value Range                         \\ \midrule
\multicolumn{1}{l|}{\multirow{2}{*}{Data Tag}} & Confidentiality (c) & 0 (Most Confidential) to 1 (Public) \\ \cline{2-3} 
\multicolumn{1}{l|}{}                          & Integrity (i)       & 0 (Lowest) to 1 (Highest)           \\ \hline
\multicolumn{1}{l|}{Code Tag}                  & Integrity (p)       & 0 (Lowest) to 1 (Highest)           \\ \bottomrule
\end{tabular}
\end{table}

The tag system used by \SysName\ is inspired by previous rule-based PIDS~\cite{morse,hossain2017sleuth}.
The focus of this paper is not on the design of the tag system.
In essence, any new type of tag can be integrated with \SysName as long as it satisfies the following conditions: 1) The tags are numerical; 2) The tags are updated through arithmetic calculation;
3) Alarms are triggered based on the tag values.
For example, we can define an additional ``exploitability" tag for all web service nodes and propose the corresponding propagation rules and alarm-triggering rules.
We leave tag designing as a practical problem in specific scenarios and use the tags mentioned above to show the improvement purely brought by \SysName's learning module.


We then introduce three adaptive parameters, which are the core of the fine-grained rule-based PIDS:
\begin{list}{\labelitemi}{\leftmargin=3pt}
 \setlength{\itemsep}{3pt}
 \setlength{\itemindent}{9pt}
    \item \textbf{Tag Initialization Parameter $(A)$} determines the initial tags of system entities.
    \item \textbf{Tag Propagation Rate Parameter $(G)$} adjusts the propagation effect of system events on tags.
    \item \textbf{Alarm Generation Threshold Parameter $(T)$} makes adjustments to the alarm generation rules.
\end{list}
We will elucidate how \SysName\ uses the three parameters to achieve more flexible rule-based detection.

\noindent\textbf{Tag Initialization:}
Proper tag initialization rules play a crucial role since they determine the initial states of system entities.
For instance, if a socket associated with a benign IP address is assigned low integrity, it could lead to many false alarms. 
However, it is relatively less discussed in previous works \cite{morse,hossain2017sleuth}.
They usually assign the initial tags based on domain knowledge without further adjustments.
Although \MORSE\ admits tag initialization policies could be learned from the previous tracing data, it did not discuss it in detail or propose a practical methodology.

We initialize the tags on the nodes when they are initially added to the provenance graph.
As illustrated in Fig. \ref{fig:overall-framework} step \normalsize{\textcircled{\scriptsize{1}}}, the detection system first checks for new system entities in the event upon receiving the system audit logs. If new entities are identified, they are added to the provenance graph and assigned initial data tags. 
Please note that when adding a new process node, we do not specifically initialize its code tag and data tag.
A process node's code tag value is inherited from the data tag of the files loaded by the process, and its data tag is propagated from its parents.
For non-process nodes, we define the adaptive parameter $A$ at the node level to facilitate fine-grained tag initialization rules. 
Specifically, for any node $n_i\in N$ in the provenance graph, $A$ stores its initial data tag with respect to its node feature. 
The node features depend on the information granularity of audit logs, such as file path, process name, command line, IP address, port, etc.
For example, if the node feature for a socket is its IP address, then all sockets with the same IP address will be assigned the same initial data tag.

\SysName's learning module refines $A$ through the training process, which will be elaborated upon in \S\ref{sec:parameter_learning}.
The training process starts from the default value $A_0$.
Since we train \SysName on the benign data, we set $A_0$ conservatively.
\NEW{In this paper, ``conservative" means the assumption of the worst-case.
Under conservative settings, we aim to capture true attacks as much as possible without considering excessive false alarms.
Rule-based PIDS such as \MORSE\ has to use the conservative setting to capture all potential attacks~\cite{morse}, while \SysName only uses the conservative setting as the initial state for training on benign data.}
\OLD{In other words, we want to ensure all potential attacks can be captured with our initial parameter $A_0$.}
For example, we assign all IP addresses an initial integrity of 0 so we do not miss any attackers entering through network communications.

Although the confidentiality scores could be customized in \SysName, we find it hard to learn them from the false alarms on the benign training data.
Instead, it often requires specific domain knowledge or personal, subjective judgments to determine what files are sensitive.
Therefore, we manually set the initial confidentiality scores and do not adjust them with the learning module.


\noindent\textbf{Tag Propagation:}
As shown in Fig. \ref{fig:overall-framework} step \normalsize{\textcircled{\scriptsize{2}}}, once the tags are initialized, specific system events will trigger the propagation of these tags along the direction of the information flow, resulting in changes to code and data tags.
In \SysName, tag propagation will pass and accumulate malicious intentions \NEW{through tag values.
For example, the \texttt{(Firefox, read, IP)} event updates the data tag of \texttt{Firefox} to the lower value between the data tags of \texttt{Firefox} and \texttt{IP}.
This ensures that potential malicious intentions carried in the information flow are preserved.}
We designed the propagation rules based on existing work~\cite{hossain2017sleuth, morse} as detailed in Table~\ref{table:propagation_rules}.







Unfortunately, such a tag propagation mechanism can suffer from the dependence explosion issue~\cite{morse}, leading to excessive alarms. 
For example, as shown in~\ref{sec:motivation}, \texttt{(bash, write, /home/user/.bash\_history)} is a common event that appears in many propagation chains of false alarms, which means this event causes large-scale maliciousness propagation on the provenance graph.
However, our investigation shows this is a commonly seen benign activity.

To mitigate such a problem, we introduce another adaptive parameter $G$, to regulate the propagation rate.
Specifically, for any edge $e\in E$ identified with (\textit{src\_node\_feature, event\_type, dest\_node\_feature}) in the provenance graph, $G$ stores its propagation rate parameter $g_e \in [0, 1]$.
Given $g_e$, the source node $src$, and the destination node $dest$, we update $tag_{dest}$ as follows during tag propagation.
\begin{equation}\label{eqn:tag_calculation}
    tag_{dest}^{new} = g_e \cdot tag_{rule} + (1-g_e) \cdot tag_{dest}
\end{equation}
Where $tag_{rule}$ is the tag value given by the propagation rules defined in Table~\ref{table:propagation_rules}. 
Usually, it equals to $tag_{src}$.
If $g_e$ is close to 0, the corresponding tag propagation will lead to small changes in the tag values and vice versa.
In this way, $G$ allows \SysName to fine-tune the propagation rules at the edge level. 

\SysName refines $G$ through the training process in the learning module, which will be elaborated upon in \S\ref{sec:parameter_learning}.
We also conservatively establish the default values of $G$ ($G_0$) to ensure all true alarms can be captured before the training starts.
Particularly, we set $g_e = 1$ for all edges $e\in E$ to make all propagation fully effective. 

\noindent\textbf{Alarm Generation:}
The alarm generation rules determine whether an alarm should be triggered. 
As shown in Fig. \ref{fig:overall-framework} step \normalsize{\textcircled{\scriptsize{3}}}, whenever an event happens, we will assess whether it satisfies the criteria to trigger alarms.
For example, if a process with a data integrity value smaller than 0.5 writes to a normal file, a file corruption alarm will be triggered.
As shown in Table~\ref{table:alarm_rules}, we generate alarms based on the event type, the subject and object tags, and the threshold.
The threshold could be fine-tuned to control the number of alarms.
For example, if event \texttt{(sshd, execve, bash)} triggers many false alarms during training, it suggests that we should decrease the alarm-triggering threshold for this event.

We introduce an adaptive parameter $T$ to enable fine-grained adjustment of the alarm threshold. 
Specifically, for any edge $e\in E$ identified with (\textit{src\_node\_feature, event\_type, dest\_node\_feature}), $T$ stores its alarm threshold $thr_e\in [0, 1]$. 
For an edge $e$, the detection function $f(e)$ is defined as follows.
\begin{equation}\label{eqn:alarm_judgement}
  f(e) = tag - thr_e =
    \begin{cases}
      malicious, & \text{if }f(e)<0 \\
      benign, & \text{otherwise}\\
    \end{cases}       
\end{equation}
where $tag$ is the relevant tag on the node of interest on edge $e_i$.
A low threshold prevents the system from generating the alarms, while a high threshold encourages the system to generate the alarms. In this way, $T$ allows \SysName to adjust the alarm-triggering rules at the edge level.

Like $A$ and $G$, \SysName's learning module refines $T$ can be refined through the training process (\S\ref{sec:parameter_learning}).
Before the training, we set the default thresholds ($T_0$) neutrally. In other words, we set $thr_i = 0.5$ for every edge $e_i$ so that we do not encourage or suppress all alarms.

Please note that \SysName's methodology is independent of specific rules or tags, making it broadly applicable to various rule-based PIDS, especially those using taint-analysis methods like \SLEUTH~\cite{hossain2017sleuth}, \HOLMES~\cite{holmes}, and \CONAN~\cite{conan}.

\subsection{Learning Module}\label{sec:parameter_learning}
With the three adaptive parameters ($A$, $G$, and $T$) we defined in the previous section, the problem of adaptative configuration for the detector is then converted to the optimization problem of the multi-variable function aiming to find the optimal values for $A$, $G$, and $T$.
In this section, we present our efforts to solve this optimization problem with the gradient descent algorithm, including defining the objective function and constraints, calculating gradients, and searching for the optimal values.

\subsubsection{Loss Function}
The first step of solving the optimization problem is to define the objective function, i.e., the loss function in the context of learning tasks.
The learning module aims to find the parameters that can reduce false alarms while maintaining the sensitivity to the true malicious events. 
Therefore, the loss function comprises two terms: the term of false alarms and the regularizer term.

The false alarm term mainly focuses on penalizing erroneously triggered alarms.
Specifically, an event $e$ triggers a false alarm means $f(e)$ should be greater than 0 but it does not.
Since $f(e) \in [-1, 1]$, we can use the Mean Squared Error $(y_e-f(e))^2$ as the loss function for all events that trigger a false alarm.
$y_e$ is set as $1$ for those benign events, and $-1$ for the malicious events.
Because we only use benign data for training, all $y_e$ should be $1$.
Please also note that we do not calculate the loss for the correctly classified events, i.e., we do not make $f(e)$ close to $1$ if $f(e)$ is already greater than $0$.
This is because the events that do not cause alarms are significantly more than the alarm-triggering events, considering them, therefore, is inefficient and would make the detection system insensible to the malicious events.
Thus, the false alarm term in the loss function can be formalized as
\begin{equation}\label{eqn:loss}
    \mathcal{L}(e) = max(0, (1-f(e))^2-1)
\end{equation}


Next, we introduce the second component: the regularizer term.
As we mentioned, \SysName is trained on benign data to learn the normal behaviors in the detection environment.
This is due to the fact that benign behaviors on a system are more consistent than attack activities.
Another reason is the malicious training data is much harder to acquire.
However, a big challenge is how to guarantee that the detection capability would not be affected if there is no malicious sample in the dataset, i.e. we would not be too lenient to capture the true alarm when reducing false alarms.

To capture all potential malicious events, the adaptive parameters are configured conservatively before training.
During training, we only want to make small adjustments to a small portion of the parameters according to the false alarms on training data.
For the rest parameters, we would like to keep them as conservative as possible so that the sensitivity to the maliciousness is not compromised.
In the area of machine learning, One-class classification (OCC) algorithms are proposed to deal with the situation where only one class of samples is available in the training set~\cite{khan2014one}.
Inspired by~\cite{tax1999data, tax1999support}, we add a regularizer term in the loss function to avoid the parameters becoming too lenient when the malicious training sample is absent.
The essential thought is to make the adaptive parameters as close to the default values ($A_0,G_0,T_0$) as possible.
By minimizing the $l_2$ distance between the adaptive parameters and the default values, we protect the detection capability during training.
Therefore, the detection loss can be formalized as
\begin{equation}\label{eqn:total_loss}
Loss = \sum_{e \in E} \mathcal{L}(e) + \alpha ||A-A_0||_2 + \gamma ||G-G_0||_2 + \tau ||T-T_0||_2
\end{equation}
where, $\alpha$, $\gamma$, and $\tau$ are the regularizer coefficients.
\NEW{In the evaluation section, we discuss the effect of the regularizer coefficients (\S\ref{sec:poisoning_attack}) and methods used to fine-tune them (\S\ref{sec:exp-setup}).}

\subsubsection{Differentiable Detection Framework}\label{subsec:differentiable}
One of the fundamental steps of using gradient descent algorithms is calculating the gradient of each variable.
To adjust the adaptive parameters based on the loss function, we need to design a differentiable detection framework, which allows us to keep the gradients of the adaptive parameters with respect to $Loss$.


According to the chain rule, for the parameter $a_n$, we have
\begin{small}
\begin{equation} \label{eq:2}
    \frac{\partial Loss}{\partial a_n} = \sum_{e \in E} \frac{\partial \mathcal{L}(e)}{\partial f} \cdot \frac{\partial f}{\partial a_n} + \alpha \cdot (a_n-a_0)
\end{equation}
\end{small}
The equations for $g_e$ and $thr_e$ \NEW{(presented in Appendix \S\ref{sec:full-proof})} hold similarly to Eq.~\ref{eq:2}.
The most challenging part of building an adaptative rule-based PIDS is to compute the gradients, i.e., $\partial f \slash \partial a_n$, $\partial f \slash \partial g_e$, and $\partial f \slash \partial thr_e$.
This is because the rule-based detection process is usually modeled as the branch selection based on rules, rather than a differentiable function.
To the best of our knowledge, no previous work has formalized or recorded the gradients of parameters in a rule-based PIDS.
In \SysName, we model the detection process as a differentiable function and calculate the gradients of each adaptive parameter.
When the tags are propagated, the corresponding gradients are updated and recorded.
In so doing, we associate the adaptive parameters with the loss, making it possible to perform the gradient descent algorithm to find the optimal parameters that minimize the loss.

As per Eq.~\ref{eqn:alarm_judgement}, $\partial f \slash \partial thr_e = -1$.
This means every time when we want to increase the value of $f$ (the direction of benign), we have to decrease the value of $thr_e$, and vice versa.
This aligns with our intuition because a lower tag value means more maliciousness in our system.
Therefore, if we want to be more lenient in the detection, we should set a lower threshold.

The formalization and calculation of $\partial f \slash \partial a_n$ and $\partial f \slash \partial g_e$ are more complicated.
According to Eq.~\ref{eqn:alarm_judgement}, calculating the gradients of the detection function $f$ is essentially computing the gradients with respect to the variable $tag$ for each node.

We start from the gradients of $a_n$.
According to the propagation policies shown in Table~\ref{table:propagation_rules}, the updated tag value $tag_{rule}$ is either the lower tag values of involved nodes $min(tag_{src}, tag_{dest})$ or a constant value $c$.
Note the subscripts \textit{src} and \textit{dest} denote the propagation direction.
If $tag_{rule} = c$, then
\begin{equation}
    \frac{\partial tag_{dest}^{new}}{\partial a_n} = (1-g_e) \frac{\partial tag_{dest}}{\partial a_n}
\end{equation}
If $tag_{rule} = tag_{src}$, then
\begin{align}\label{equation:alpha_grads}
    \frac{\partial tag_{dest}^{new}}{\partial a_n} &= g_e\frac{\partial tag_{src}}{\partial a_n} + (1-g_e) \frac{\partial tag_{dest}}{\partial a_n}
\end{align}


Since $a_n$ is defined as the initial value of the tag on node $n$, the initial gradient of $a_n$ is
\begin{equation}\label{eq:init_an_grads}
    \frac{\partial tag_{n'}}{\partial a_n} = 
    \begin{cases}
        1, & \text{if } n' = n \\
        0, & \text{if } n' \neq n
    \end{cases}
\end{equation}

Then we focus on calculating gradients of $g_e$, which is more complicated.
From Eq.~\ref{eqn:tag_calculation}, we know that $tag_{dest}^{new}$ is determined by $tag_{dest}$, $tag_{rule}$, and $g_e$.
And $tag_{dest}$, $tag_{rule}$ are related to the propagation rates of previous events.
Therefore, $tag_{dest}^{new}$ is influenced by the propagation rates of the current event (denoted by $g_e$) and all previous events (denoted by $g_{e'}$) happened on $src$ and $dest$ nodes.
Consequently, for each propagation, we must update $\partial tag_{dest}^{new}/\partial g_e$ and $\partial tag_{dest}^{new}/\partial g_{e'}$ simultaneously.
The case of $g_{e'}$ is similar to that of $a_n$, we have
\begin{align}\label{equation:gamma_grads_1}
    \frac{\partial tag_{dest}^{new}}{\partial g_{e'}} &= g_e\frac{\partial tag_{rule}}{\partial g_{e'}} + (1-g_e) \frac{\partial tag_{dest}}{\partial g_{e'}}
\end{align}
Calculating gradients of $g_e$ involves the Product Rule in calculus, which is
\begin{small}
\begin{align}\label{equation:gamma_grads_2}
    \frac{\partial tag_{dest}^{new}}{\partial g_e} &= g_e\frac{\partial tag_{rule}}{\partial g_e} + (1-g_e) \frac{\partial tag_{dest}}{\partial g_e}+tag_{rule}-tag_{dest}
\end{align}
\end{small}
The full proof of Eq.~\ref{equation:gamma_grads_1} and Eq.~\ref{equation:gamma_grads_2} can be found in the Appendix \S\ref{sec:full-proof}.


As previously stated, $g_e$ are set according to the event features \textit{(src\_node\_feature,event\_type,dest\_node\_feature)}.
Therefore, before the propagation happens, 
\begin{equation}\label{equation:initial_lambda_grads}
    \frac{\partial tag_n}{\partial g_e} = 0, \forall\ e \in E, \forall\ n \in N
\end{equation}
We use Eq.~\ref{equation:initial_lambda_grads} to initialize the gradient of $g_e$ and update them according to Eq.~\ref{equation:gamma_grads_1} and ~\ref{equation:gamma_grads_2}.
\OLD{Besides, similar to the case of $a_n$, when the tags are not updated (i.e., $M=tag_{dest}$), we do not have to update the gradients on the related nodes.}

In summary, to employ the gradient descent algorithm, we calculate the gradients of the adaptive parameters with respect to the loss.
We first calculate the initial gradient for each node.
Afterward, the gradients are updated according to our equations when the tags are propagated.
\NEW{Please note that unlike GNN models such as GCN~\cite{kipf2016semi} and GraphSage~\cite{hamilton2017inductive}, which only consider the n-hop neighbor of a node, \SysName\ fine-tunes every edge and node in the graph that can influence the detection results by propagating their gradients.}

\subsubsection{Training and Testing}
After calculating the gradients of the adaptive parameters with respect to the loss using the differentiable detection framework, we can now utilize the gradient descent algorithm to optimize the parameters.
In machine learning, this process is referred to as “training”.
The learned parameters are stored as a customized configuration, which is then used to set up \SysName prior to testing.

As shown in Fig.~\ref{fig:overall-framework}, an epoch in training comprises the forward and backward propagation.
Before training starts, all adaptive parameters, $A$, $G$, and $T$, are configured as the default settings $A_0$, $G_0$, and $T_0$.
Instead of random initialization\cite{chen2019gradient}, we use the most conservative setting as the starting point to keep the sensitivity to the maliciousness during training.

When \SysName processes the events, tags are propagated among the graph, updating the gradients according to Eq.~\ref{equation:alpha_grads}, ~\ref{equation:gamma_grads_1} and ~\ref{equation:gamma_grads_2}, and generate the detection results.
Next, we calculate the loss and back-propagate the gradients of loss according to Eq.~\ref{eq:2}.
Finally, parameters are updated in the opposite direction of the gradient as follows:
\begin{equation}
    p_{new} = p_{old} - l \cdot \frac{\partial Loss}{\partial p_{old}}
\end{equation}
where $p$ is the adaptive parameters (which could be $a_n$, $g_e$, or $thr_e$), $p_{new}$ is the new parameters, $p_{old}$ is the old one, $\frac{\partial Loss}{\partial p_{old}}$ is the gradient, and $l$ is the learning rate.
We repeat the training process when the maximum epoch is reached or the changes in the result become sufficiently small.
Before testing, we configure the parameters according to what we learned in the training stage.
Then, \SysName can process audit data and conduct detection as introduced in \S~\ref{sec:overalldesign}.

%% file: sections/Implementation.tex
The entire system (including data parsing, tag initializing, tag propagating, alarm generating, and training/testing framework) consists of 5KLoC of Python.
We implement the differentiable detection framework by creating two dictionaries for each tag to store the gradients with respect to $A$ and $G$.
As for $T$, because its gradients are unrelated to the node tags, there is no need to store them within the node.
The space used to store those gradient dictionaries during training is analyzed in
\S~\ref{subsubsec: runtime-overhead} and \S~\ref{appendix: gradients_num}


%% file: sections/Evaluation.tex

Our evaluation aims to answer the following five research questions: 
1) How effectively can \SysName\ detect the attacks, especially in terms of reducing false alarms?
2) How efficient is \SysName\ compared with the SOTA PIDS in terms of detection latency and runtime overhead? (\S\ref{subsec:evaluation_overhead})
3) How robust is \SysName\ against adversarial attacks such as mimicry attacks and data poisoning attacks? (\S\ref{subsec:adversarial_attack})
4) How do different components affect the training outcome and the detection performance of \SysName? (\S\ref{subsec: ablation-study})
5) Can \SysName acquire explainable knowledge via our learning module? (\S\ref{subsec:case-study})

\subsection{Experiment Settings}
\subsubsection{Datasets}



We evaluate \SysName\ using public forensic datasets from the DARPA Transparent Computing program and datasets generated within simulated environments in collaboration with an SOC.

\noindent \textbf{DARPA Datasets.} The DARPA Transparent Computing program was organized between 2016 and 2019 to perform several red team assessments.
In two weeks, the data collecting teams deploy collectors on several target hosts~\cite{noauthor_transparent_nodate}.
We use the public-available datasets from Engagement 3 (E3) and Engagement 5 (E5) in our evaluation.
\cite{engagement3data,engagement5data} provides a detailed description of attacks performed in relevant DARPA datasets.

\noindent\textbf{Simulated Environments.}
Moreover, we collaborate with an industry SOC to acquire additional datasets within realistically simulated scenarios to avoid the problem of “close-world data”~\cite{li2023nodlink}.
Specifically, the SOC furnishes detailed host setups from real-world operating environments.
Subsequently, we simulate APT attacks~\cite{noauthor_kimsuky_nodate} employing the Atomic Red Team~\cite{noauthor_atomicredteam_nodate} and online malware repositories~\cite{chaos}.
The simulated scenarios encompass five APT attacks across three distinct real-world operational settings, elaborated in detail in \S\ref{appendix:attack-scenarios}. 


\noindent\textbf{Data Labeling.}
For each attack scenario, we label entities and events on the kill chains as malicious according to the attack reports.
Although previous work~\cite{wang2022threatrace, rehman2024flash} provided entity-level data labels, they did not specify the data labeling strategy or context.
We also observed that the amount of malicious labels in their dataset is excessively large (e.g., over 12 thousand system entities were marked as malicious within a 30-hour period on the \CADETS\ from DARPA Engagement 3), which is impractical for a real-world SOC.
Thus, we decided to use our data labels in the evaluation for a fair and unbiased comparison between \SysName and the baseline systems.
We also make our data labels publicly available to facilitate future research.

\subsubsection{Experiment Setup}\label{sec:exp-setup}
We deployed \SysName and performed all experiments on an Ubuntu 22.04.3 Linux Server with an Intel(R) Xeon(R) Platinum 8358 CPU @ 2.60GHz and 1.0 TB memory.
We partition each dataset into training and testing sets, adhering to the assumption stated in \S\ref{subsec:threatmodel} that the training set should not contain any malicious activities. 
Specifically, we find the starting time of the first attack.
Then, all data produced before that date becomes the training set, whereas the remaining data becomes the testing set.

To avoid biased results due to overfitting, we performed cross-validation when evaluating the performance of \SysName.
However, standard \textit{k}-fold cross-validation is unsuitable for streaming logs because the time series cannot be freely split into \textit{k} groups, as the subsequent tags depend on the previous ones.
Therefore, we utilized time series cross-validation, which preserves the chronological order of data.
We set a time window and trained the parameters using the data within this window. Afterward, we moved the time window forward until creating \textit{k} different training sets.
In our experiment, we set \textit{k}=3 and the length of the time windows to be around 2/3 of the total length of the training set.
We also discuss overfitting in \S~\ref{sec:overfitting_discussion}.

Like many other machine learning systems, \SysName\ also relies on appropriate hyperparameters, especially three regularizer coefficients $\alpha, \gamma, \tau$, to train the accurate and robust model (the effect of these hyperparameters are shown in \S\ref{sec:poisoning_attack}).
We provide two methods to fine-tune the hyperparameters, depending on whether a validation set is available.
If a validation set exists, we perform grid searching, a generic hyperparameter tuning approach.
We train multiple models with different hyperparameter combinations and evaluate their performance on the validation set.
The default conservative detector is run on the validation set as the baseline.
We select the hyperparameters from the model with the fewest false alarms and equivalent true alarms to the baseline.
For instance, the optimal hyperparameters of E3-\CADETS\ is $\alpha=0.1, \gamma=0.1, \tau=0.1$ and the learning rate is 0.01.
However, a validation set is usually not available for a PIDS in practice.
Even if it exists, it may not encompass all possible attacks.
In this case, we propose a heuristic-based method to fine-tune the regularizer coefficients.
Recall that the training starts from the conservative adaptive parameters, false alarms in the training set will loosen the conservative settings.
The regularizer coefficients are added to the loss function to avoid being too lenient and missing the true alarms.
Therefore, the values of the regularizer coefficients $\alpha$, $\gamma$, and $\tau$ are determined by the following question: how many false alarms caused by a node/edge can we tolerate in the training set to avoid being overly lenient?
Before training starts, we set a number for the allowed false alarms ($N$), and we can estimate an approximate value of the regularizer coefficients based on $N$:
$\alpha \approx 3N, \gamma \approx 3N, \tau \approx 12N$ (see Appendix \S\ref{sec:full-proof-regularizer} for the mathematical proof).
Please note that these are approximate estimations depending on the extent to which we trust the ``benign" training data.
A practical way might be: set regularizer coefficients using heuristics; if missing alarms occur compared to the conservative baseline, multiply the coefficients by a factor (e.g., 10) and retrain.
Conversely, if no alarms are missed for a long time $T$, reduce the coefficients by a factor and retrain.

\subsubsection{Baseline Detectors}
We compare \SysName\ with five SOTA PIDS: \FLASH~\cite{rehman2024flash}, \KAIROS~\cite{cheng2023kairos}, \SHADEWATCHER~\cite{zengy2022shadewatcher}, \NODLINK~\cite{li2023nodlink}, and \MORSE~\cite{morse} to evaluate their performance from different perspectives.
We chose \FLASH, \KAIROS, and \NODLINK\ because they are the SOTA embedding-based PIDS.
In addition, their implementations are open-sourced, allowing us to test them on different datasets.
Since some other embedding-based PIDS~\cite{wang2022threatrace, han2020unicorn, holmes} have already been evaluated in these works and the result shows that \FLASH, \KAIROS, and \NODLINK\ outperform them, we didn't include them in our evaluation.
We chose \MORSE\ since it is the SOTA rule-based PIDS, and we want to evaluate the improvements from our differentiable adaptation framework in \SysName.
To get a fair and unbiased comparison, we use the settings from \MORSE\ in the evaluation.
According to~\cite{cheng2023kairos,rehman2024flash} and our communication with the authors, the detection systems of \SHADEWATCHER\ and \PROGRAPHER\ are not fully open-source due to proprietary license restrictions.
Although the data preprocessing module of \SHADEWATCHER\ is open-sourced, we tried our best but could not locate the preprocessing output files used for the following training and testing.
We then realized that it was not feasible for us to perfectly replicate their systems for an unbiased comparison.
Therefore, we used the code to evaluate the efficiency and latency of data preprocessing and the detection results reported in their paper to evaluate detection accuracy.
For \FLASH, \KAIROS, and \NODLINK, we used their open-sourced code as the basis for the evaluation.
We reimplemented \MORSE\ according to their paper.

\subsection{Detection Accuracy}
\label{subsec:effectiveness}
\subsubsection{False Alarm Events Reduction}
We first focus on the reduction of false alarm events, which requires the detection granularity at the event level.
We compare \SysName\ with the SOTA event-level PIDS \SHADEWATCHER\ and the classical rule-based PIDS \MORSE.
Due to the close-source nature of \SHADEWATCHER, we used the reported detection results in their paper.
The results show that both \MORSE\ and \SysName successfully detected all attacks in the testing dataset.
But \SysName\ reduces the false alarm rate by over 93\% (15.66x) compared to \SHADEWATCHER\ and over 95\% (20.45x) compared to \MORSE.
Additionally, \SHADEWATCHER\ holds
80\% of events for training and only 10\% for testing, while the testing set of \SysName\ is over seven times larger than the testing set of \SHADEWATCHER.
This shows that \SysName\ does not require as much data as \SHADEWATCHER\ for training.

\begin{table}[h!]\footnotesize
\centering
\caption{Comparison with the baselines on the \TRACE\ dataset from Engagement 3 in terms of false alarm events}
\begin{tabular}{c|c|c|c}
\toprule
                 & \MORSE\ & \SHADEWATCHER\ & \SysName\  \\ \midrule
                 
\# of Consumed Events       &    5,188,230   & 724,236      & 5,188,230 \\ \hline
\# of False Alarm Events  &    22,500   & 2,405         & 1,099      \\ \hline
False Alarm Rate &   0.434\%    & 0.332\%      & 0.0212\% \\
\bottomrule
\end{tabular}
\label{table:fp-event-comparison}
\vspace{-1em}
\end{table}

One advantage of the rule-based PIDS like \SysName\ and \MORSE\ is that it can offer semantic-rich alarms, while \SHADEWATCHER\ only flags deviations from patterns observed during training without explaining the alarms.
We then compared the number of false alarms generated by \SysName\ and \MORSE\ into different categories, shown in Table~\ref{table:morse-comparison}.

\begin{table*}[h!]\scriptsize
\centering
\caption{Comparison with the baseline on different datasets regarding false alarms. T-3, T-5, C-3, and C-5 represent the \TRACE\ and \CADETS\ datasets from Engagement 3 \& 5. S-1 to S-3 denote Cloud, Streaming, and Dev datasets from the SOC.}
\begin{tabular}{c|cc|cc|cc|cc|cc|cc|ccr}
\toprule
\multirow{2}{*}{Datasets} &
  \multicolumn{2}{c|}{FileExe} &
  \multicolumn{2}{c|}{MemExec} &
  \multicolumn{2}{c|}{ChPerm} &
  \multicolumn{2}{c|}{Corrupt} &
  \multicolumn{2}{c|}{DataLeak} &
  \multicolumn{2}{c|}{Escalate} &
  \multicolumn{3}{c}{Total False Alarms} \\ \cmidrule(l){2-16} 
 &
  \multicolumn{1}{c|}{Base} &
  Ours &
  \multicolumn{1}{c|}{Base} &
  Ours &
  \multicolumn{1}{c|}{Base} &
  Ours &
  \multicolumn{1}{c|}{Base} &
  Ours &
  \multicolumn{1}{c|}{Base} &
  Ours &
  \multicolumn{1}{c|}{Base} &
  Ours &
  \multicolumn{1}{c|}{Base} &
  \multicolumn{1}{c|}{Ours} &
  \multicolumn{1}{c}{Reduction} \\ \midrule
T-3 &
  \multicolumn{1}{c|}{9} &
  0 &
  \multicolumn{1}{c|}{{49.3K}} &
  {3.76K} &
  \multicolumn{1}{c|}{1} &
  1 &
  \multicolumn{1}{c|}{{626}} &
  2 &
  \multicolumn{1}{c|}{{955}} &
  {14} &
  \multicolumn{1}{c|}{{673}} &
  0 &
  \multicolumn{1}{c|}{{52.1K}} &
  \multicolumn{1}{c|}{{3.78K}} &
  \textbf{{13.78x}} \\ \hline
C-3 &
  \multicolumn{1}{c|}{41} &
  14 &
  \multicolumn{1}{c|}{N/A} &
  N/A &
  \multicolumn{1}{c|}{{6}} &
  {6} &
  \multicolumn{1}{c|}{{13.2K}} &
  {272} &
  \multicolumn{1}{c|}{{96}} &
  {50} &
  \multicolumn{1}{c|}{81} &
  {1} &
  \multicolumn{1}{c|}{13.4K} &
  \multicolumn{1}{c|}{{343}} &
  \textbf{{39.07x}} \\ \hline
T-5 &
  \multicolumn{1}{c|}{181} &
  170 &
  \multicolumn{1}{c|}{403K} &
  79.8K &
  \multicolumn{1}{c|}{0} &
  0 &
  \multicolumn{1}{c|}{34.4K} &
  19.9K &
  \multicolumn{1}{c|}{26.9K} &
  1.20K &
  \multicolumn{1}{c|}{18.4K} &
  28 &
  \multicolumn{1}{c|}{483K} &
  \multicolumn{1}{c|}{101K} &
  \textbf{4.78x} \\ \hline
C-5 &
  \multicolumn{1}{c|}{{1.63K}} &
  {0} &
  \multicolumn{1}{c|}{N/A} &
  N/A &
  \multicolumn{1}{c|}{N/A} &
  N/A &
  \multicolumn{1}{c|}{{1.81M}} &
  {6.64K} &
  \multicolumn{1}{c|}{{7.87K}} &
  {2} &
  \multicolumn{1}{c|}{{1.42K}} &
  {0} &
  \multicolumn{1}{c|}{{1.83M}} &
  \multicolumn{1}{c|}{{6.64K}} &
  \textbf{{276x}} \\ \hline
S-1 &
  \multicolumn{1}{c|}{3.16K} &
  0 &
  \multicolumn{1}{c|}{N/A} &
  N/A &
  \multicolumn{1}{c|}{26} &
  0 &
  \multicolumn{1}{c|}{100} &
  73 &
  \multicolumn{1}{c|}{22} &
  21 &
  \multicolumn{1}{c|}{N/A} &
  N/A &
  \multicolumn{1}{c|}{3.31K} &
  \multicolumn{1}{c|}{94} &
  \textbf{35.21x} \\ \hline
S-2 &
  \multicolumn{1}{c|}{177} &
  53 &
  \multicolumn{1}{c|}{N/A} &
  N/A &
  \multicolumn{1}{c|}{0} &
  0 &
  \multicolumn{1}{c|}{12} &
  9 &
  \multicolumn{1}{c|}{14} &
  7 &
  \multicolumn{1}{c|}{N/A} &
  N/A &
  \multicolumn{1}{c|}{203} &
  \multicolumn{1}{c|}{69} &
 \textbf{2.94x} \\ \hline
S-3 &
  \multicolumn{1}{c|}{18} & 0
   &
  \multicolumn{1}{c|}{N/A} &
  N/A &
  \multicolumn{1}{c|}{0} &
  0 &
  \multicolumn{1}{c|}{29} &
  16 &
  \multicolumn{1}{c|}{23} &
  22 &
  \multicolumn{1}{c|}{N/A} &
  N/A &
  \multicolumn{1}{c|}{60} &
  \multicolumn{1}{c|}{38} &
  \textbf{1.58x} \\
  \bottomrule
\end{tabular}
\label{table:morse-comparison}
\vspace{-1.5em}
\end{table*}


Table~\ref{table:morse-comparison} shows that \SysName\ can reduce false alarms by over 90\% (11.49x) on average for all datasets compared with the non-adaptive \MORSE.
It outperforms \MORSE\ in every alarm category.
Those remaining false alarms, especially in C-5, S-2, and S-3, cannot be removed for the following reasons.
First, some “false” alarms are not purely benign. They are related to some (potential) attack nodes (e.g., configuration files in \texttt{\footnotesize /tmp/atScript/atomic-red-team-Gray\_dev1.0/*} in S-2) but cannot be mapped to specific attack steps directly, thus not labeled as “malicious.”
Second, some nodes and edges do not exist or trigger any alarm in the training set.
Third, since our approach aims to be as conservative as possible, some parameters are tuned just enough to eliminate false alarms in the training set.
However, these conservative settings may still cause false alarms when applied to the testing set.

\begin{table}[h!]\scriptsize
\centering
\caption{Comparison with the baselines in terms of node-level detection accuracy}
\begin{tabular}{c|c|c|c|c}
\toprule
             & TP & FP(0-hop) & FP(1-hop) & FN(0-hop) \\\midrule
\multicolumn{5}{c}{Engagement3 \CADETS}                         \\\midrule
\FLASH\        &    16    & 4503   & 4485 &  10  \\
\KAIROS\       &    15    & 1017    & 1003  &  11  \\
\NODLINK$^1$\      &    3   & 120   & 114 &  2 \\
\NEW{\MORSE}      &    \NEW{16}   & \NEW{51}    & \NEW{43}  &  \NEW{10} \\
\SysName\      &    16   & 34    & 26  &  10 \\\midrule
\multicolumn{5}{c}{Engagement3 \TRACE}                        \\\midrule
\FLASH\       &    5    & 27202   & 27178 &  19 \\
\NODLINK$^1$\ &    4    & 170   &  170  &  0  \\
\NEW{\MORSE}      &    \NEW{10}   & \NEW{243}    & \NEW{234}  &  \NEW{14} \\
\SysName\     &    10    & 12       & 11     &  14 \\\midrule
\multicolumn{5}{c}{Engagement3 \THEIA}                        \\\midrule
\FLASH\       &    2    & 53230   & 53050 &  13 \\
\KAIROS\      &    12    & 3566     &  3422  &  3 \\
\NODLINK$^1$\ &    4    & 62     & 58   &  0 \\
\NEW{\MORSE}      &    \NEW{11}   & \NEW{220}    & \NEW{213}  &  \NEW{4} \\
\SysName\     &    11    & 194     & 187   &  4 \\
\bottomrule
\end{tabular}
\\ \footnotesize \raggedright
$^1$Since \NODLINK\ only provides detected process, we evaluate it on process detection accuracy.
\label{tab:node-accuracy-comparison}
\end{table}

\subsubsection{False Alarm Entity Reduction}
While \SysName\ operates as an event-level detector, we still compare \SysName\ with the SOTA entity-level PIDS \FLASH, \KAIROS, and \NODLINK(\KAIROS\ and \NODLINK\ give the result at graph-level, but they also support entity-level detection).
We transformed the alarm events triggered by \MORSE\ and \SysName\ into alarm entities using the following method: an entity is considered an alarm entity if it is involved in any alarm events.

It is noteworthy that only a subset of each DARPA dataset is utilized for training and testing in~\cite{wang2022threatrace, rehman2024flash, cheng2023kairos}.
To simulate the real scenarios in SOC, we utilize the entire dataset for evaluation, dividing the training and testing sets as described in~\S\ref{sec:exp-setup}.
We retrained the model and performed detection using their code.
Additionally, we noticed \FLASH\ and \THREATRACE\ did not count the FPs within the two-hop distance from the labeled attack entities in the ground truth while counting the TPs within the two-hop distance from the detected entities~\cite{cheng2023kairos}. 
To ensure a fair comparison without providing excessive leniency, we report the TP and FP results, as well as the one-hop FP result, in which we exclude the FP entities within a one-hop distance from the ground truth.
We also did not count the attack entities in the 2-hop neighborhood of the detected entities as TP like~\cite{rehman2024flash,cheng2023kairos,wang2022threatrace} did.
These strict experiment settings and data labeling methods can explain the difference between our results and the results in their papers, but we believe it is fair and necessary to assess the performance of PIDS in real-world scenarios.

We evaluated the detection accuracy on \CADETS, \TRACE, and \THEIA\ from Engagement 3 since those datasets were commonly used by the baselines.
Table~\ref{tab:node-accuracy-comparison} illustrates the comparison among \SysName and the baselines.
On all datasets, \SysName\ demonstrates superior performance in identifying more TPs while maintaining fewer FPs. 
\FLASH\ can report a fair amount of TPs but produce an excessive amount of FPs.
It leverages GNN to learn $k$-hop neighborhood structures. 
While this technique shows promising performance when there is a significant anomaly within $k$ hops, it could degrade when the training set is limited~\cite{rehman2024flash}.
\KAIROS\ can filter out FPs with the Anomalous Time Window Queue~\cite{cheng2023kairos}.
Although it guaranteed decent detection accuracy at the time-window level; the entity-level accuracy, however, is not satisfactory.
\NODLINK\ detects most attack processes.
However, missing other relevant entities, such as files and network sockets, requires additional expert efforts to investigate the reported alarms.
\MORSE\ and \SysName achieve a lower false positive rate in node-level detection.
This is attributed to the semantic-rich alarms that provide additional information for alarm filtering.
For instance, a file corruption alarm involves two entities: the process and the corrupted (benign) files.
As the corrupted files should not be reported as malicious entities, they can be easily filtered out.
This explains why \MORSE\ triggers thousands of false alarms at the event level, but only hundreds of nodes are incorrectly alarmed.
For embedding-based PIDS, there is no such semantics for alarm filtering.
We include the filtered results in Table~\ref{tab:node-accuracy-comparison} as we believe it highlights an advantage of rule-based PIDS.


\SysName successfully detects all attacks in the dataset.
As an event-level detector, \SysName does not report entities that are not directly related to the attack events.
For example, if a malware is downloaded and executed, \SysName reports the MalFileCreation alarm and the FileExec alarm, which reveals the malware process, file, and parent process.
We do not immediately report the entry network node used by attackers to avoid alarm fatigue, which explains the FNs of \SysName.
The investigations on the relevant nodes can be conducted after the first response.

\subsection{Efficiency}\label{subsec:evaluation_overhead}
In previous sections, we highlighted the simplicity of \SysName's rule-based detection framework compared to other embedding-based PIDS.
In this section, we compare \SysName\ with the SOTA PIDS \FLASH, \KAIROS, \NODLINK, \SHADEWATCHER\footnote{Our evaluation was based solely on the open-sourced portion of \SHADEWATCHER.}, and \MORSE\ to evaluate their efficiency by running their code on our test environment.
\subsubsection{Detection Latency}
The detection latency analysis encompasses three dimensions: \textbf{buffer time}, \textbf{preprocessing time}, and \textbf{detection time}.
We have already introduced buffer time in \ref{subsec:background-PIDS}.
Preprocessing time refers to the duration taken to convert the raw audit logs into a data structure that the detection systems can process.
It involves data parsing, data cleaning, noise reduction, preprocessing, feature extraction, and so on.
Detection time refers to the interval between the completion of data processing and the moment the result is produced.
It is noteworthy that since the detection granularity is different in different PIDS, the detection time might not reflect the actual latency of each system.
For instance, a detector based on the whole graph may take longer to deliver detection results of each graph compared with a detector on the entity/event level.
However, this does not necessarily mean the former is slower because a large graph could contain many entities and events.
Therefore, we calculate the total time spent on detection across the entire dataset.

We evaluated the latency on \TRACE\ and \CADETS\ from DARPA Engagement 3 since they are covered in the experiments of all baselines.
Our evaluation focuses on the testing stage since training can be conducted offline.
We modified their code to obtain unbiased results.
For example, \FLASH\ divides the data processing into two stages in their code and uses files to store intermediate results. 
And \KAIROS\ employs the PostgreSQL database and stores the intermediate results in files.
These are the steps not required in a real-time streaming pipeline.
Consequently, we exclude the I/O time to focus solely on measuring the ``pure" preprocessing and detection time.

The result is shown in Table~\ref{tab:latency-comparison}.
Unlike other PIDS that embeds the graphs and thus require a buffer time ranging from several minutes to hours, \SysName processes audit logs in a streaming fashion, eliminating the need for any buffer time.
\SysName is also faster in preprocessing because the logs are not preprocessed for the machine-learning model.
Moreover, the tags of \SysName\ are much simpler than the state vectors used by \FLASH\ and \KAIROS.
By avoiding the use of text embedding or GNNs to aggregate semantic and contextual graph information, \SysName\ achieves detection speeds over 10 times faster than the baseline methods.
Lastly, compared to the rule-based PIDS \MORSE, \SysName\ only takes a few seconds longer to detect logs spanning 4 to 7 days, once again demonstrating the superiority of rule-based PIDS in terms of latency.


\begin{table}[h]\scriptsize
\centering
\caption{Comparison of detection latency}
\begin{tabular}{c|c|c|c}
\toprule
             & Buffer Time & Preprocessing Time & Detection Time \\\midrule
\multicolumn{4}{c}{Engagement3 \TRACE}                         \\\midrule
\FLASH\        &       57:49      & 107:50          & 64:24          \\
\SHADEWATCHER\ &     N/A$^1$        & 100:22          &      3:40$^2$          \\
\NODLINK\       & 00:10      &  135:42               &   2:48             \\
\NEW{\MORSE}      & \NEW{0}           & \NEW{58:20}           & \NEW{1:29}           \\
\SysName\      & 0           & 58:20           & \NEW{1:31}           \\\midrule
\multicolumn{4}{c}{Engagement3 \CADETS}                        \\\midrule
\KAIROS\       & 15:00      &  15:34               &   29:46             \\
\FLASH\       & 82:52      &  18:57               &   7:41             \\
\NODLINK\       & 00:10      &  6:18               &   6:41             \\
\NEW{\MORSE}       & \NEW{0}           & \NEW{7:22}           & \NEW{1:19}             \\
\SysName\      & 0           & 7:22           & 1:23          \\
\bottomrule
\end{tabular}
\\ \footnotesize \raggedright
$^1$We did not find a clear number in their codes or paper.\\
$^2$\SHADEWATCHER\ extracts the last 10\% interactions as the testing set, while the testing set of us is around 2.5 times larger.
\label{tab:latency-comparison}
\vspace{-2em}
\end{table}

\subsubsection{Runtime Overhead}\label{subsubsec: runtime-overhead}
Another important metric reflecting the efficiency of a PIDS is the runtime overhead to the system.
In this section, we evaluate the runtime overhead of \SysName\ during detection and conduct a comparative study with other SOTA PIDS.
We also analyze the memory consumption of \SysName\ during the training stage.

\begin{figure}[htbp]
  \centering
  \vspace{-1em}

  \begin{subfigure}[b]{0.45\textwidth}
  \centering
  \includegraphics[width=0.85\textwidth]{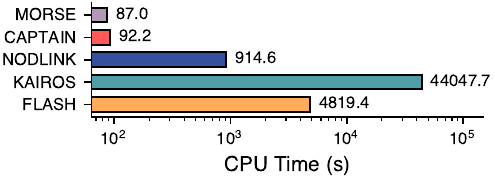}
  \vspace{-0.5em}
    \caption{Total CPU time used}
    \label{fig:resource_cpu}
  \end{subfigure}
  \hfill
  \begin{subfigure}[b]{0.45\textwidth}
  \centering
    \includegraphics[width=0.85\textwidth]{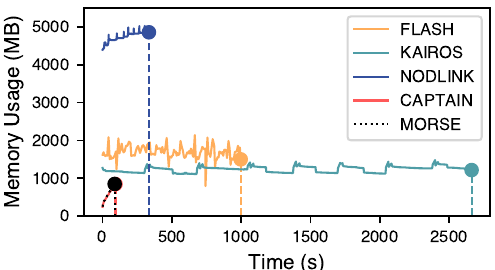}
    \vspace{-0.5em}
    \caption{Memory usage over time}
    \label{fig:resource_memory}
  \end{subfigure}

  \caption{Comparison of resource consumption when detecting on DARPA Engagement 3 \CADETS. \NEW{In Fig.~\ref{fig:resource_memory}, the memory usage curve of \SysName\ and \MORSE\ are overlapped, showing their similar efficiency performance.}}
  \label{fig:main}
\end{figure}

\textbf{Runtime Overhead in Detection:}
We use the Python \texttt{resource} module to evaluate the resource consumption of detectors on CPU mode.
As admitted in~\cite{zengy2022shadewatcher}, GPUs may not be available in most real-life threat detection scenarios.
We calculate the total CPU time in user mode.
The result is shown in Fig.~\ref{fig:resource_cpu}.
Please note that the total CPU time used in Fig.~\ref{fig:resource_cpu} could be more than the detection time in Table~\ref{tab:latency-comparison} because of the multi-core CPU usage by deep learning models.
Since \SysName\ only relies on straightforward arithmetic operations on real-number tags, it requires significantly less CPU time than the embedding-based PIDS (around 2\% of \FLASH\ and around 0.02\% of \KAIROS), which need complex matrix and vector operations due to their use of neural networks.
Moreover, \SysName's fine-grained detection rules only increase CPU time by a modest 5.6\% compared to \MORSE.
Given the significant improvement on \MORSE\ in detection accuracy (over 90\% reduction in false alarms) and the substantial decrease in CPU time compared to other embedding-based PIDS, we believe the slightly higher CPU usage than \MORSE\ is acceptable in real-world detection scenarios.

We use the \texttt{psutil} library in Python to monitor the live memory usage during detection.
The comparison of memory usage over time is shown in Fig~\ref{fig:resource_memory}.
\SysName\ finishes detection much faster than the embedding-based PIDS (as already shown in Table~\ref{tab:latency-comparison}) and achieves the lowest memory usage throughout the entire detection process.
The memory usage and detection time of \SysName\ and \MORSE\ are similar, shown by the overlapping curves in Fig.~\ref{fig:resource_memory}.
The slight difference in memory usage between \SysName\ and \MORSE\ is due to the storage of fine-grained rules.
In general, \SysName\ retains the lightweight and fast characteristics of the rule-based PIDS.

\textbf{Runtime Overhead in Training:}
The training of the detectors can be performed offline, where users can allocate ample time and computing resources, including GPUs.
Therefore, we did not evaluate the runtime overhead and the time used for training.
However, as \SysName records and propagates the gradients for each tag among the big provenance graph, a reasonable concern is that saving those gradients would consume much memory when there is a dependency explosion problem.
We evaluated the number of non-zero gradients saved by \SysName during training.
The results, shown in Fig.~\ref{fig:gradient_length}, illustrate that most nodes only need to store a small number of non-zero gradients during training.
Due to the space limit, the detailed analysis can be found in \S\ref{appendix: gradients_num}.


\subsection{Resilience Against Adversarial Attacks}\label{subsec:adversarial_attack}
In this section, we discuss the robustness of \SysName\ against two main-stream adversarial attacks: adversarial mimicry attack and training set poisoning attack.

\subsubsection{Adversarial Mimicry Attack} 
\begin{figure}[h!]
\vspace{-0.5em}
    \centering
    \includegraphics[width=0.85\linewidth]{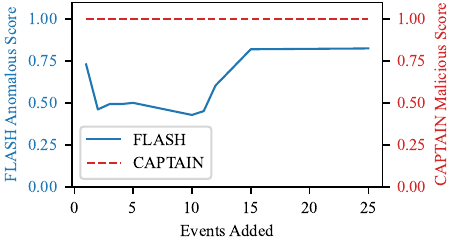}
    \caption{Adversarial mimicry attack against \SysName\ and \FLASH\ (use the attack entity \texttt{/tmp/test} as an example)}
    \label{fig:mimicry_attack}
    \vspace{-0.5em}
\end{figure}

Mimicry attacks on PIDS involve altering the provenance data and incorporating more benign features to ``mimic benign behaviors", thereby evading detection.
In this section, we evaluated the robustness of \SysName\ against mimicry attacks using the attack methodology in~\cite{goyalsometimes, rehman2024flash}, i.e. inserting benign structures into the attack graphs.
Our mimicry attack contains two steps.
First, we extract some events of benign system entities from the normal training data.
Next, we create ``fake" events by substituting the benign entities with the attack entity, simulating the attack entity performing activities similar to those of the benign entities.
To verify the effectiveness of the mimicry attack, we use \FLASH\ as the baseline.
For evaluation purposes, we use the \CADETS\ from Engagement 3 because of its relatively small scale.
The details about our mimicry experiment can be found in the Appendix \ref{appendix:mimicry_attack}.
The result is shown in Fig.~\ref{fig:mimicry_attack}.

In general, both \SysName\ and \FLASH\ showed robustness against our mimicry attack attempt to some extent.
The attack entity \texttt{/tmp/test} is still detected by both systems even after we add some normal activities.
However, we see a significant drop in the anomalous score of \FLASH\ when we introduce a relatively small number of events (fewer than 10), which is also confirmed in their paper~\cite{rehman2024flash}.
On the other hand, \SysName\ remains unaffected against the mimicry attack, demonstrating superior robustness compared to the baseline.

The robustness of \SysName\ results from three main reasons.
First of all, the propagation of \SysName is guided by heuristic-based rules, while the propagation and aggregation of \FLASH\ and other embedding-based PIDS are based on the neural network.
Second, \SysName is an event-level detector with a finer granularity than entity-level detectors.
As mentioned before, \SysName\ gives the detection result for each streaming event without any buffer time, which means any mimicry insertion after the real attack activities is useless.
For those PIDS with the buffer time, any events within the same time window and related to the same entity can affect the final detection result.
Third, \SysName is not an anomaly-based detector.
In other words, the detection is performed based on the conservative, heuristic rules, while the normal data is only used to reduce the false alarms.
Therefore, adding normal features can not compromise the detection capability as it is guaranteed by the conservative detection rules.

\subsubsection{Training Set Poisoning Attack}\label{sec:poisoning_attack} Training set poisoning refers to the malicious manipulation of the training set.
In the context of cyber security and intrusion detection, it usually involves polluting the benign training set with adversarial events/entities.
When the detector is trained on a ``benign" dataset that has been polluted, it learns patterns or features introduced by attackers.
These patterns and features, once learned, could assist attackers in evading detection in future instances.


We add the regularizer term to the loss function to control the sensitivity of \SysName\ to the training set.
This enhances \SysName's robustness against dataset poisoning attacks.
In this section, we present a real case of Pine Backdoor \& Phishing Email attack from Engagement 3 \TRACE.
In this attack, a vulnerable \texttt{pine} connected to the email server \texttt{128.55.12.73}, downloaded and executed an email attachment.
However, IP \texttt{128.55.12.73} had multiple activities and caused some “false” alarms in the training set.
Please note that this violates our assumption stated in \S~\ref{subsec:threatmodel} that the training set must not be compromised by attack entities or events.
Nevertheless, this makes it a perfect example to test the robustness against data poisoning attacks.

We analyze the critical parameters for detecting this attack: the initial integrity tag ($a$) of IP node \texttt{128.55.12.73}, to illustrate the effect of the regularization term.
Without the regularization term in the loss function, \texttt{128.55.12.73} will be assigned a high integrity score (near 1.0) due to its activities in the training set, which lets us miss the phishing email attack starting from it.
However, with the help of the regularizer terms ($\alpha=10$), \SysName\ is more cautious during training.
Like the ``grey node" motivating example in \S\ref{sec:motivation}, \SysName\ assigns the integrity score as the mediocre 0.488, which removes many false alarms caused by this IP while successfully capturing the phishing email attack during testing (more details can be found in \S\ref{subsec:case-study}).
The propagation rate of event \texttt{(pine, read, 128.55.12.73)} is also more conservative against the polluted training set due to the regularizer term.

\subsection{Ablation Study}
\label{subsec: ablation-study}
\subsubsection{Individual Adaptive Parameters}
We conduct an ablation study on each adaptive parameter.
As each parameter can be learned independently, we have seven separate experiments, each optimizing a subset of the parameters, i.e. $\{A\}$, $\{G\}$, $\{T\}$, $\{A,G\}$, $\{A,T\}$, $\{G,T\}$, and $\{A,T,G\}$.

\begin{figure}[h!]
\vspace{-0.5em}
    \centering
    \includegraphics[width=0.9\linewidth]{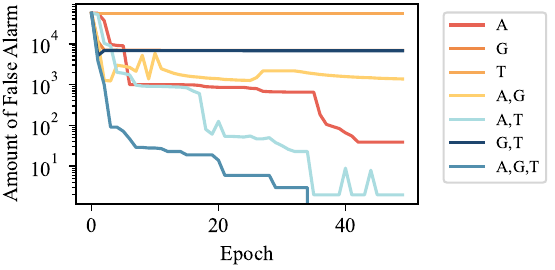}
    \vspace{-0.5em}
    \caption{Ablation study on each parameter.}
    \label{fig:ablation_test}
    \vspace{-0.5em}
\end{figure}

According to Fig.\ref{fig:ablation_test}, adjusting the threshold ($T$) alone may not have much effect on the detection results because without changing the initial tags and the propagation rates, the updated tags still tend to be 0 or 1.
Fig.\ref{fig:ablation_test} also shows that tuning all these three parameters can accelerate the training process.
Although $\{A,T\}$ and $\{A,T,G\}$ achieve similar good detection results on the testing set, $\{A,T,G\}$ converges using fewer epochs during training.
It should be noted that tuning $A$ plays the most major role in mitigating false alarms among these three adaptive parameters.
In certain datasets (S-3 and C-3), adjusting $A$ alone sometimes results in fewer false alarms than tuning parameters $A$, $G$, and $T$ together because, without the benefit of learning propagation rates and thresholds, we have to assign higher integrity scores to more nodes to counteract false alarms during training.
However, reducing false alarms more crudely carries the risk of being excessively lenient.

\subsubsection{Learning Rate in Optimization}\label{sec:learning_rate_study}
We conducted an experiment on different learning rates on the DARPA Engagement 3 \CADETS.
The detailed results can be found in \S~\ref{subsec:ablation-lr}.
As illustrated in Fig.~\ref{fig:lr_exp}, a higher learning rate usually leads to a sharper loss decline during the initial training epochs.
However, for ongoing training, a lower learning rate may yield a finer-tuned search process and lower loss, necessitating a balance between training speed and training quality.

\subsection{Case Study}
\label{subsec:case-study}
To illustrate the interpretability of \SysName's detection process, we delve into the training outcomes of the E3-\TRACE\ dataset as a case study.
These cases also show the improvement of \SysName compared to existing rule-based PIDS.

\subsubsection{Better Detection Capability due to Fine-Grained Rules}
The most significant improvement of \SysName\ is its fine-grained detection capability, enabled by the adaptive parameters.
As mentioned in \S~\ref{sec:motivation}, unlike the discrete, human-assigned trustworthiness levels and universal rules used in~\cite{morse,hossain2017sleuth,holmes}, the fine-tuned numerical parameters of each entity and event in the graph can better help distinguish nuances between benign and malicious patterns.

\begin{figure}[htbp]
  \centering
  \vspace{-1em}

  \begin{subfigure}[b]{0.5\textwidth}
  \centering
  \includegraphics[width=0.9\textwidth]{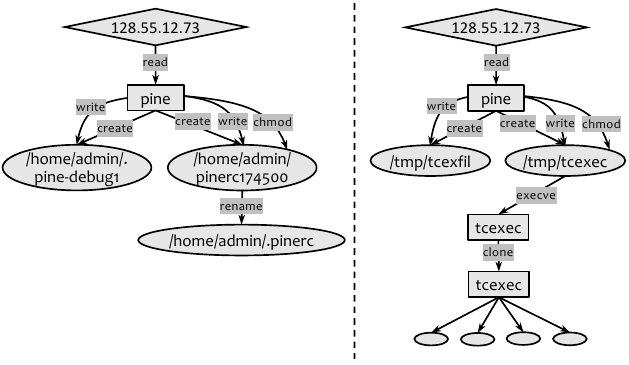}
  \vspace{-0.5em}
    \caption{The graph on the left side represents normal behavior in the training set, while the graph on the right side depicts an attacker executing a malicious email attachment.}
    \label{fig:case_study:alpha_tau}
  \end{subfigure}
  \hfill
  \begin{subfigure}[b]{0.5\textwidth}
  \centering
    \includegraphics[width=0.9\textwidth]{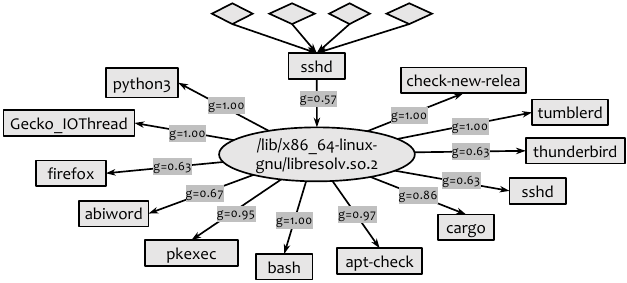}
    \vspace{-0.2em}
    \caption{\NEW{\SysName\ assigns different propagation rates ($g_e$) to events to control dependency explosion more precisely.}}
    \label{fig:case_study:gamma}
  \end{subfigure}

  \caption{\NEW{Case study from Engagement 3 \TRACE. We omitted some irrelevant details to keep the graphs clean and succinct.}}
  \label{fig:case_study_detection_ability}
\end{figure}

Fig.~\ref{fig:case_study:alpha_tau} demonstrates how \SysName\ outperforms the baselines in distinguishing between two similar behaviors using fine-tuned parameters.
In both cases, process \texttt{pine} reads from \texttt{128.55.12.73} (possibly the email server) and creates, writes, and changes the permissions of a new file.
It is difficult for previous rule-based PIDS~\cite{hossain2017sleuth, morse} to assign a tag for \texttt{128.55.12.73}.
Take the SOTA \MORSE\ as an example -
if \MORSE\ initialize the node \texttt{128.55.12.73} as a benign node with the data integrity score as 1.0, it would miss at least the first stage of the attack.
On the other hand, if \MORSE\ assigns the initial tag as ``Untrusted" (this is what they did in order to capture all attacks), it would trigger multiple false ``Malicious File Creation" and ``Change Permissions" alarms on the benign graph.
Instead, \SysName\ assigns a moderate initial tag $a_n$, 0.488, to this IP and slightly tunes down the threshold $thr$ for the event \texttt{(pine, create, /home/admin/pinerc174500)} from 0.5 to 0.478.
As $0.478<0.488$, the creation of the file will not be alarmed, but in real attacks, the $thr$ for the event \texttt{(pine, create, /tmp/tcexec)} is still 0.5, meaning creating a malicious file would still trigger alarms ($0.5>0.488$).


\SysName's fine-grained rules also help address the dependency explosion issue caused by \texttt{libresolv.so.2}.
Unlike \MORSE, which sets universal decay/attenuation factors for all processes, \SysName\ learns distinct propagation rates for different events.
As shown in Fig.~\ref{fig:case_study:gamma}, \SysName\ assigns lower propagation rates to events that frequently cause dependency explosions (e.g., being read by \texttt{sshd} or \texttt{firefox}) while maintaining high propagation rates for other events.
Even for the same process, \texttt{firefox}, \SysName\ learns to suppress the dependency propagation when \textit{reading} this file ($g_e=0.63$) and remains sensitive to maliciousness propagation when \textit{loading} this file ($g_e=0.99$).
Consequently, \SysName\ avoids being overly lenient for potential attack events when addressing the dependency explosion issue.

The fine-grained parameters of \SysName\ also outperform many embedding-based PIDS in detecting mimicry attacks.
For instance, in Fig.~\ref{fig:case_study:alpha_tau}, the attacker can modify the file name of \texttt{/tmp/tcexec} with the knowledge that \texttt{pine} might interact with a file like \texttt{/home/admin/pinerc+numbers}.
Since generalization techniques like Word2Vec and Sentence2Vec are widely used in embedding-based PIDS, they can manipulate the file name by appending random digits to \texttt{pinerc}, making it similar to \texttt{pinerc174500} after embedding.
However, to evade \SysName, the attacker must know the exact file name of \texttt{pinerc174500}.
In addition, deep learning usually relies on the amount of training data.
But there are only four events related to file \texttt{/home/admin/pinerc174500} during the 7-day training set, making it challenging for the neural networks to learn such patterns.

\subsubsection{Robustness brought by Differentiable Tag-Propagation Framework}
Some previous rule-based PIDS~\cite{nodoze} can customize their detection system using benign training data.
However, the adjustments are based on the training set rather than the detection results.
For example, \NODOZE\ assigns a distinct anomaly score to each event based on its frequency in the training set.
However, due to training set poisoning and living-off-the-land attack techniques, the frequent events in the training set cannot be fully trusted.
For instance, the events related to \texttt{dbus-daemon} are very common in the training set of \TRACE\ in Engagement 3.
Typical events such as \texttt{dbus-daemon} writes or reads \texttt{/var/run/dbus/system\_bus\_socket} occur over 100,000 times during seven days.
Such events and the \texttt{dbus-daemon} entity would be assigned with a low anomaly score in frequency-based systems like \NODOZE.
However, the low anomaly scores allow the attackers to deceive the detectors using living-off-the-land techniques related to the D-Bus process~\cite{dbusattack}.
In contrast, \SysName\ does not change the parameters related to \texttt{dbus-daemon} because, despite the large number of events associated with \texttt{dbus-daemon} in the training set, no false alarms are triggered by it.
\SysName\ adjusts the parameters related to 
an event based on the false alarms caused by it rather than the frequency of it.

\OLD{\noindent\textbf{Tag Initialization (A) Adjustments:}
Examining the parameters in $A$, the adjustments from $A_0$ primarily focused on network objects, showcasing a broad spectrum of modifications.
Notably, internal IP addresses, such as \texttt{128.55.12.10, 128.55.12.122} and \texttt{128.55.12.73}, were assigned initial integrity values close to 1.  
Other IP addresses, like \texttt{65.214.39.18} and \texttt{216.9.245.101}, are assigned initial integrity values around $0.5$. 
Additionally, certain IP addresses, such as \texttt{208.17.90.11} and \texttt{162.93.202.80}, were endowed with initial integrity values slightly exceeding 0.
The discrepancies in the learned integrity values reflect the varying levels of trust \SysName\ places in these IP addresses.
Internal IP addresses, in particular, triggered numerous alarms through potentially malicious behaviors during the training period. To reduce such false alarms, \SysName\ learned to trust these internal IP addresses, leading to the assignment of elevated integrity values.
Other IP addresses, appearing less frequently and engaging in less sensitive activities, received neutral integrity values. 
Moreover, some IP addresses, despite participating in sporadic sensitive activities eventually deemed benign, did not garner significant trust from \SysName. Consequently, their initial integrity values underwent a minimal change from $A_0$.
Compared with manual whitelisting, \SysName\ can assign integrity with finer granularity, allowing it to navigate the trade-offs between true positives and false positives more effectively.}

\OLD{\noindent\textbf{Propagation Rate (G) Adjustments:}
Examining the parameters in $G$, the adjustments from $G_0$ predominantly focus on read and write operations on files.
Notably, events involving \texttt{bash} reading from \texttt{.bash\_history} received a relatively low propagation rate, approximately $0.5$, while other events associated with \texttt{.bash\_history} maintained a high propagation rate close to 1. 
As discussed in \S~\ref{sec:motivation}, \texttt{bash} frequently engages in reading and writing activities with \texttt{.bash\_history} during routine tasks, potentially causing dependency explosions and triggering numerous false alarms. Through the training process, \SysName\ learned to mitigate false positives by reducing the propagation rates for events involving \texttt{bash} reading from \texttt{.bash\_history}. Meanwhile, \SysName\ remained cautious with other \texttt{.bash\_history} related events, maintaining relatively high propagation rates to ensure malicious intent can still propagate through alternative pathways.
The fine-grained adjustment of propagation rates by \SysName\ diminishes false positives while preserving its robustness, thereby preventing attackers from easily evading detection.
}

\OLD{\noindent\textbf{Alarm Generation Threshold (T) adjustments:}
Upon scrutinizing the parameters in $T$, the adjustments encompass various events capable of triggering alarms.
Notably, we observed that the threshold for generating data leak alarms is set close to 0 for processes such as \texttt{sshd}, \texttt{thunderbird}, and \texttt{firefox} when they send messages to internal IP addresses.
In contrast, the threshold for generating data leak alarms when these processes send messages to other IP addresses remains around the default value of $0.5$.
Furthermore, the internal IP addresses associated with these processes are distinct. For instance, \texttt{sshd} exhibits a low alarm generation threshold exclusively when sending messages to \texttt{128.55.12.122} and \texttt{128.55.12.10}, while \texttt{thunderbird} displays a low threshold when communicating with \texttt{128.55.12.73}.
These observations indicate that \SysName\ has effectively learned to diminish false alarms by lowering the threshold for generating such alarms at the edge level. These adjustments are finely tuned to specific processes communicating with particular IP addresses, while leaving the alarm generation threshold unaltered for other processes communicating with these IP addresses or for these processes communicating with other IP addresses.
This fine-grained and nuanced approach not only allows \SysName\ to effectively reduce false alarms but also fortifies it against living-off-the-land attacks that seek to emulate benign activities. The resilience to such attacks persists unless the attackers can successfully gain access to the internal IP addresses featured in the training dataset, thereby significantly heightening the complexity and challenge of executing such an attack.
}


%% file: sections/Discussion.tex
\subsection{Overfitting in Parameter Learning}\label{sec:overfitting_discussion}
Overfitting is an important issue in machine learning.
As \SysName learns the benign patterns in the one-class manner starting from conservative settings, overfitting is, however, a preferred and safer strategy.
In one-class learning, we want to generate a tightest boundary for the target class (benign patterns).
Therefore, \SysName avoids generalizing excessively to minimize the risk of evasion, as demonstrated in our example in \S\ref{subsec:case-study}.

\subsection{More Optimization Algorithms and Online Learning}\label{sec:more_learning_algo}
In this paper, we utilize the gradient descent algorithm for optimization due to its simplicity and clearness.
We can reach optimal values more quickly and accurately, using algorithms such as \textit{Nesterov Accelerated Gradient (NAG)}~\cite{nesterov1983method}, \textit{Adadelta}~\cite{zeiler2012adadelta}, \textit{RMSprop}, \textit{Adagrad}~\cite{duchi2011adaptive}, \textit{Adam}~\cite{kingma2014adam}, \textit{AdaMax}, \textit{Nadam}, and so on.
These algorithms require the first-order gradient, offered by the differentiable detection framework in \SysName.
We leave the implementation of these algorithms as the extensions of \SysName.

Long-term maintenance is important in real-world scenarios.
Although we assess \SysName in an offline setting, \SysName's learning module can compute loss upon receiving the feedback for each event.
Subsequently, the learning module can promptly update the adaptive parameters for subsequent detection.

%% file: sections/Conclusion.tex
This paper introduces \SysName, a rule-based PIDS capable of automatically adapting to detection environments, enhanced by three adaptive parameters: tag initialization ($A$), propagation rate ($G$), and threshold ($T$).
The differentiable detection framework enables the optimization using the gradient descent algorithm.
To our knowledge, this is the first effort to incorporate gradient descent methods in optimizing rule-based PIDS.
We evaluated \SysName on several datasets.
The results demonstrate the superior detection capability, significantly reducing false alarms, detection latency, and runtime overhead, outperforming the SOTA baselines.

%% file: sections/Appendix.tex
\section{Attack Scenarios in Datasets}
\label{appendix:attack-scenarios}

For benign scenarios, the SOC collaborating with us provided three detailed operation environments in the real world: Streaming simulates a data store and streaming server; Dev simulates a development server; Cloud simulates a cloud server.
The simulated APT attack chains are generated using the Atomic Red Team \cite{noauthor_atomicredteam_nodate}.
We designed three types of simulated attacks: fully randomized, partially randomized, and deterministic attacks.
Each step in a fully randomized attack is randomly selected from the Atomic Red Team library.
A partially randomized attack has determined steps at some stages but randomly chooses steps for the rest of the steps.
A deterministic attack has defined steps at every attack stage.
We generated one deterministic and three partially randomized attacks in our evaluation. 
We conduct the experiments following the attack campaign pattern in DARPA Engagements to make it as realistic as possible. 
Table \ref{scenario-detail} specifies the apps or processes running in each simulated environment.

\begin{table}[h!]\scriptsize
\caption{Simulated Scenario Summary}
\centering
\begin{tabular}{cp{0.67\linewidth}}
\toprule
Attack     & Attack Description \\ 
\midrule
 ReverseShell(RS) &   Connect to the victim's host, collect system information, and install multiple applications  \\
 WebShell(WS)           &   Connect to the victim's host, collect system information and modify system configurations  \\
 AttackChain(AC)           &   A randomly generated attack chain using Atomic Red Team~\cite{noauthor_atomicredteam_nodate}  \\
 Kimsuky(Kim)          &   A simulated North Korean APT Kimsuky \\
 Chaos        &   A malicious payload that allows remote control \\

\toprule

 Benign     & App or Process Involved \\ 
\midrule

   Streaming   &  Kafka, Mysql, Nginx, Redis, Zookeeper \\
 Developing          &   Iptables, Zabbix, Gitlab, VSCode, Influxdb, lvextend, redis-server, Qingtengyun, Baota, docker\\
 Cloud          &   finalshell, postgres, web.py, Apache Struts 2, saltstack, Cloud Workload Protection Platforms\\
\bottomrule
\end{tabular}
\label{scenario-detail}
\vspace{-2em}
\end{table}

\section{\SysName Policies}
\label{appendix:policies}
We adopt tag propagation rules, detailed in Table~\ref{table:propagation_rules}, from \MORSE~\cite{morse} for unbiased comparison.
$dtag$ refers to the data tag and $ptag$ refers to the code tag.
Please refer to the original paper for formal and detailed descriptions of the hyperparameters such as $T_{qb}$, $T_{qe}$, $d_b$, and $d_e$.
All hyperparameters are set according to the recommendation in~\cite{morse}.
We also adopt the alarm generation rules from \MORSE~\cite{morse}, detailed in Table~\ref{table:alarm_rules}.
$incl\_exec(p)$ means $p$ is contains the execution permission.
$socket(o)$ holds when $o$ refers to a socket.

\begin{table*}[htbp]\scriptsize
\centering
\caption{Tag propagation policies}
\begin{tabular}{@{}c|c|c|c|c@{}}
\toprule
\multirow{2}{*}{Event}          & \multirow{2}{*}{Tag to update} & \multicolumn{3}{c}{New tag value for different subject types} \\
\cline{3-5}
               &               &           benign         & suspect       & suspect environment  \\ 
\midrule
$create(s, x)$   & \( x.dtag \)  & \( s.dtag \)            & \( s.dtag \)  & \( s.dtag \)        \\
\hline
$read(s, x)$     & \( s.dtag \)  & \( \min(s.dtag, x.dtag) \) & \( \min(s.dtag, x.dtag) \) & \( \min(s.dtag, x.dtag) \) \\
\hline
$write(s, x)$    & \( x.dtag \)  & \( \min(s.dtag + a_b, x.dtag) \) & \( \min(s.dtag, x.dtag) \) & \( \min(s.dtag + a_e, x.dtag) \) \\
\hline
\multirow{2}{*}{$load(s, x)$} & $s.ptag$ & \multicolumn{3}{c}{\( \min(s.ptag, x.itag) \)} \\
\cline{2-5}
 & $s.dtag$ & \multicolumn{3}{c}{\( \min(s.dtag, x.dtag) \)} \\
\hline
\multirow{2}{*}{$exec(s, x)$} & $s.ptag$ & \( x.itag \) & \( \min(x.itag, susp\_env) \) & \( x.itag \) \\
\cline{2-5}
 & $s.dtag$ & \( \langle 1.0, 1.0 \rangle \) & \( \min(s.dtag, x.dtag) \) & \( \min(s.dtag, x.dtag) \) \\
\hline
\multirow{2}{*}{$inject(s, s')$} & $s'.stag$ & \multicolumn{3}{c}{ \( \min(s'.stag, s.itag) \)} \\
\cline{2-5}
 & $s'.dtag$ & \multicolumn{3}{c}{\( \min(s.dtag, s'.dtag) \)} \\
\hline
$periodically$:  & \( s.dtag \)  & \( \max(s.dtag, d_b * s.dtag + (1 - d_b) * T_qb) \) & no change & \( \max(s.dtag, d_e * s.dtag + (1 - d_e) * T_qe) \) \\
\bottomrule
\end{tabular}

\label{table:propagation_rules}
\end{table*}



\begin{table*}[htbp]\scriptsize
\centering
\caption{Alarm generation policies}
\begin{tabular}{c|l|c|c|c}
\toprule
\textbf{Name} & \textbf{Description} & \textbf{Operation(s)} & \textbf{Data integrity condition} & \textbf{Other conditions} \\ \midrule
MemExec & Prepare binary code for execution & $mmap(s,p)$, $mprotect(s,p)$ & $s.itag < 0.5$ & $incl\_exec(p)$ \\ \hline
FileExec & Execute file-based malware & $exec(s,o)$, $load(s,o)$ & $s.itag < 0.5$ & $s.ptag > 0.5$ \\ \hline
Inject & Process injection & $inject(s,s')$ & $s.itag < 0.5$ & $s'.ptag > 0.5$ \\ \hline
ChPerm & Prepare malware file for execution & $chmod(s,o,p)$ & $s.itag < 0.5$ & $incl\_exec(p)$ \\ \hline
Corrupt & Corrupt files & $write(s,o)$, $mv(s,o)$, $rm(s,o)$ & $s.itag < 0.5$ & - \\ \hline
Escalate & Privilege escalation & $any(s)$ & $s.itag < 0.5$ & changed userid \\ \hline
DataLeak & Confidential data leak & $write(s,o)$ & $s.itag < 0.5$ & $s.ctag < 0.5$, $socket(o)$ \\ \hline
MalFileCreation & Ingress Tool Transfer & $create(s,o)$ & $s.itag < 0.5$ & $File(o)$ \\
\bottomrule
\end{tabular}

\label{table:alarm_rules}
\end{table*}

\section{Mathematical Proof of the Optimization}\label{sec:full-proof}
In this section, we prove the equations of gradients of $A$, $G$, and $T$ to the loss, i.e. Eq.~\ref{equation:alpha_grads}, Eq.~\ref{equation:gamma_grads_1}, and Eq.~\ref{equation:gamma_grads_2} in \S\ref{sec:parameter_learning}.
According to the loss function Eq.\ref{eqn:total_loss} and the chain rule in calculus, we have:

\begin{footnotesize}
\begin{equation}\label{equation:gradient_three_parameters}
\begin{aligned}
    \frac{\partial Loss}{\partial a_n} &= \sum_{e \in E} \frac{\partial \mathcal{L}(e)}{\partial f} \cdot \frac{\partial f}{\partial a_n} + \alpha \cdot (a_n-a_0)\\
    \frac{\partial Loss}{\partial g_e} &= \sum_{e \in E} \frac{\partial \mathcal{L}(e)}{\partial f} \cdot \frac{\partial f}{\partial g_e} + \gamma \cdot (g_e-g_0)\\
    \frac{\partial Loss}{\partial thr_e} &= \sum_{e \in E} \frac{\partial \mathcal{L}(e)}{\partial f} \cdot \frac{\partial f}{\partial thr_e} + \tau \cdot (thr_e-thr_0)
\end{aligned}
\end{equation}
\end{footnotesize}
where $\frac{\partial \mathcal{L}}{\partial f}$ is identical for all events according to Eq.~\ref{eqn:loss} and $\alpha$, $\gamma$, and $\tau$ are pre-defined hyperparameters.
We have
\begin{footnotesize}
\begin{equation*}
\begin{aligned}
    f(e) &= tag_{dest}^{new} - thr_e\\
    &= g_e \cdot tag_{rule} + (1-g_e) \cdot tag_{dest} - thr_e\\
    \frac{\partial f}{\partial thr_e} &= -1\\
    \frac{\partial f}{\partial a_n} &= g_e \cdot \frac{\partial tag_{rule}}{\partial a_n} + (1-g_e) \cdot \frac{\partial tag_{dest}}{\partial a_n}
\end{aligned}
\end{equation*}
\end{footnotesize}

$f(e)$ is influenced by the propagation rate of the current event (denoted by $g_e$) and all previous events (denoted by $g_{e'}$).
Consequently, every time when $tag_{dest}^{new}$ is updated, we recalculate $\frac{\partial f}{\partial g_e}$ and $\frac{\partial f}{\partial g_{e'}}$.

\begin{footnotesize}
\begin{equation}\label{equation:gradient_eprime}
\begin{aligned}
    Since\ \frac{\partial g_{e}}{\partial g_{e'}} &= 0,\\
    \frac{\partial f}{\partial g_{e'}} &=
        g_e \cdot \frac{\partial tag_{rule}}{\partial g_{e'}} + (1-g_e) \cdot \frac{\partial tag_{dest}}{\partial g_{e'}}
\end{aligned}
\end{equation}
\end{footnotesize}

The calculation of $\frac{\partial f}{\partial g_e}$ involves the product of two functions.
According to the Product Rule:

\begin{footnotesize}
\begin{equation*}
\begin{aligned}
    if\ h(x) = f(x)g(x),\ then\ h'(x) = f'(x)g(x) + f(x)g'(x)
\end{aligned}
\end{equation*}
\end{footnotesize}
we have
\begin{footnotesize}
\begin{equation}\label{equation:gradient_e}
\begin{aligned}
    \frac{\partial f}{\partial g_{e}} = g_e\frac{\partial tag_{rule}}{\partial g_e}+tag_{rule} + (1-g_e) \frac{\partial tag_{dest}}{\partial g_e}-tag_{dest}
\end{aligned}
\end{equation}
\end{footnotesize}


\section{Mathematical Proof of the Heuristic Hyperparameter Setting}\label{sec:full-proof-regularizer}
According to Eq.\ref{equation:gradient_three_parameters}, to let $\frac{\partial Loss}{\partial a_n}$, $\frac{\partial Loss}{\partial g_e}$, and $\frac{\partial Loss}{\partial thr_e}$ be zero, we have
\begin{footnotesize}
\begin{equation}
\begin{aligned}
    \alpha &= -\sum_{e \in E} \frac{\partial \mathcal{L}(e)}{\partial f} \cdot \frac{\partial f}{\partial a_n} \cdot (a_n-a_0)^{-1}\\
    \gamma &= -\sum_{e \in E} \frac{\partial \mathcal{L}(e)}{\partial f} \cdot \frac{\partial f}{\partial g_e}\cdot (g_e-g_0)^{-1}\\
    \tau &= -\sum_{e \in E} \frac{\partial \mathcal{L}(e)}{\partial f} \cdot \frac{\partial f}{\partial thr_e}\cdot (thr_e-thr_0)^{-1}
\end{aligned}
\end{equation}
\end{footnotesize}
Since $\mathcal{L}(e) = max(0, (1-f(e))^2-1)$ and $-1<f(e)<0$,
\begin{footnotesize}
\begin{equation*}
\begin{aligned}
    -\sum_{e \in E} \frac{\partial \mathcal{L}(e)}{\partial f} = \sum_{e \in E} 2(1-f(e))
\end{aligned}
\end{equation*}
\end{footnotesize}

$\frac{\partial f}{\partial thr_e} = -1$, $0 < \frac{\partial f}{\partial a_n} < 1$, $-1 < \frac{\partial f}{\partial g_e} < 0$.
As mentioned in \S\ref{sec:detection_design}, $a_0$ is usually 0 while $g_0$ is 1.
We can make the following estimations:

\begin{footnotesize}
\begin{equation}\label{eqn:estimation_basis}
\begin{aligned}
    f(e) \approx -0.5,
    \partial f / \partial a_n \approx 0.5,
    \partial f / \partial g_e \approx -0.5\\
    a_n-a_0 \approx 0.5,
    g_e-g_0 \approx -0.5,
    thr_e-thr_0 \approx -0.25
\end{aligned}
\end{equation}
\end{footnotesize}

Let $N(e)$ be the number of the allowed false alarms related to event $e$ and $N(n)$ be the number of the allowed false alarms related to node $n$.
\begin{footnotesize}
\begin{equation}\label{eqn:coefficient_estimation}
\begin{aligned}
    \alpha &= \sum_{e \in E} 2(1-f(e)) \cdot \frac{\partial f}{\partial a_n}\cdot (a_n-a_0)^{-1} \approx 3N(n)\\
    \gamma &= \sum_{e \in E} 2(1-f(e)) \cdot \frac{\partial f}{\partial g_e}\cdot (g_e-g_0)^{-1} \approx 3N(e)\\
    \tau &= \sum_{e \in E} 2(1-f(e)) \cdot \frac{\partial f}{\partial thr_e}\cdot (thr_e-thr_0)^{-1} \approx 12N(e)\\
\end{aligned}
\end{equation}
\end{footnotesize}


\section{Additional Experiment Results}



\subsection{Gradients Storage During Training}\label{appendix: gradients_num}
\begin{figure}[ht]
\centering

\begin{subfigure}{0.2\textwidth}
    \includegraphics[width=\linewidth]{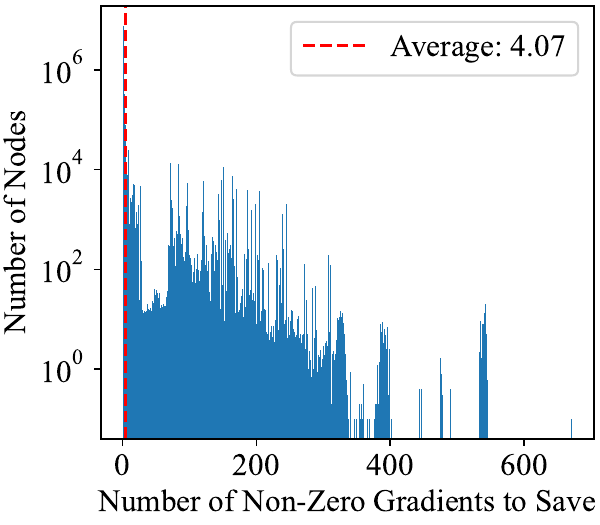}
    \caption{T-3}
    \label{fig:sub1}
\end{subfigure}
\begin{subfigure}{0.2\textwidth}
    \includegraphics[width=\linewidth]{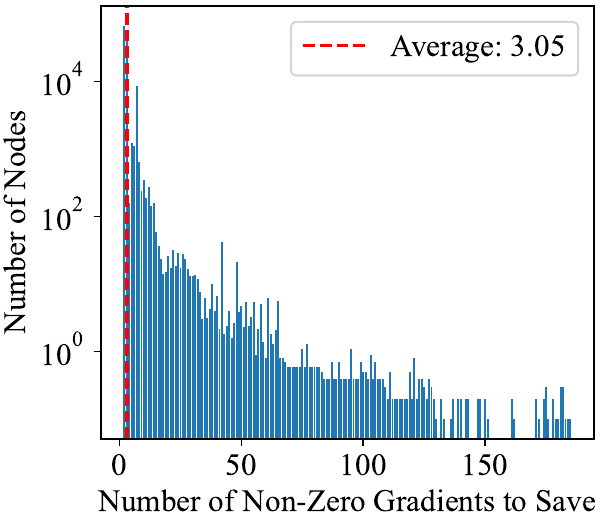}
    \caption{C-3}
    \label{fig:sub2}
\end{subfigure}

\vspace{-0.0cm} 

\begin{subfigure}{0.2\textwidth}
    \includegraphics[width=\linewidth]{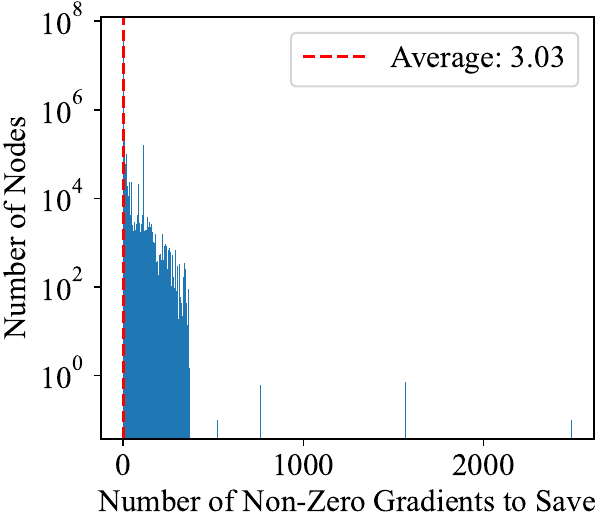}
    \caption{T-5}
    \label{fig:sub3}
\end{subfigure}
\begin{subfigure}{0.2\textwidth}
    \includegraphics[width=\linewidth]{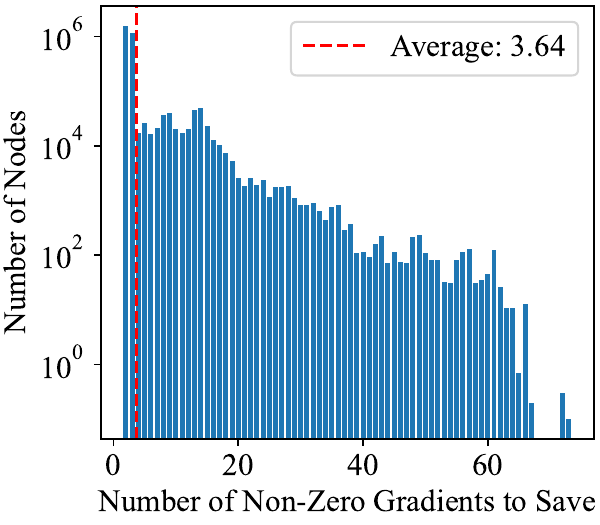}
    \caption{C-5}
    \label{fig:sub4}
\end{subfigure}

\caption{The distribution of the numbers of non-zero gradients to $A$ and $G$ that each nodes need to save during tag propagation.}
\label{fig:gradient_length}
\end{figure}

We maintain two dictionaries for each node $n$ to save the gradients of $tag_n$, $\nabla_n^A:\{n':\frac{\partial tag_{n}}{\partial a_n'},n' \in N\}$ and 
$\nabla_n^G:\{e:\frac{\partial tag_{n}}{\partial g_e},e \in E\}$.
Theoretically, the worst case for saving $\nabla_n^A$ is $O(|N|^2)$, and $O(|N||E|)$ for saving $\nabla_n^G$.
Please note that we only store non-zero gradients.
Every time after the gradients are calculated, we add the non-zero values into $\nabla_n^A$ and $\nabla_n^G$ and discard the too-small gradients (e.g., $<10^{-5}$).
In practice, most gradients would always be zero for three reasons:
1) In most cases, the provenance graph is relatively sparse.
For example, a specific process would only read a small set of files on the system;
2) The gradients only get updated when the event changes the tags.
Therefore, most benign nodes and edges would not be added to $\nabla_n^A$ and $\nabla_n^G$;
3) As $g_e$ is smaller than 1, some gradients would become near-zero after many iterations.
Fig.~\ref{fig:gradient_length} clearly illustrates the long-tailed distribution of the non-zero gradients on the DARPA dataset, which means we only need to store a tiny number of non-zero gradients for most nodes.
The average numbers are less than 5 for all those four datasets.
Hence, the additional overhead is reasonable during training.

\subsection{Mimicry Attack Experiment}
\label{appendix:mimicry_attack}
We insert events into the dataset to mimic benign behaviors involving malicious entities.
we used the E3 \CADETS\ for this experiment. 
In the attack, \texttt{/tmp/test} file is downloaded and executed, serving as a command and control malware to carry out the rest of the attack. 
We identified a benign file \texttt{/dev/tty} to serve as the target to mimic and replicated interactions with \texttt{/dev/tty} on \texttt{/tmp/test} to let \texttt{/tmp/test} mimic normal behaviors.

\subsection{Ablation Study on Learning Rate}
\label{subsec:ablation-lr}
\begin{figure}[h!]
    \centering
    \includegraphics[width=0.7\linewidth]{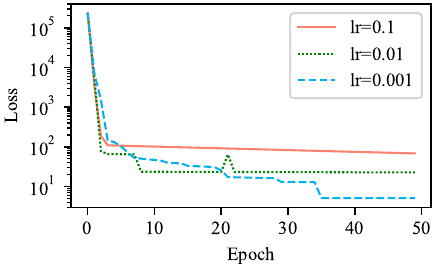}
    \caption{Training loss using different learning rates.}
    \label{fig:lr_exp}
\end{figure}
Fig.~\ref{fig:lr_exp} shows the training loss with different learning rates on \CADETS\ from DARPA Engagement 3.
From the figure, if the training is limited to 20 epochs (due to time and resource issues), a learning rate of 0.01 is advisable; for longer training durations, 0.001 is likely a more preferred choice.